# 1. Introduction

The search for the top quark and the Higgs boson, the only as yet unobserved ingredients of the Standard Model (SM), is expected to dominate the area of particle physics in the coming years. While the Higgs boson has always been naturally associated to the breaking of the electro-weak symmetry, the recent experimental discovery that the top quark is heavier than the $W$ and $Z$ has led physicists to speculate that it also may be sensitive to physics on the electro-weak scale. For this reason, the theoretical determination of its physical properties in the context of the SM and their systematic comparison with experiment may reveal an underlying, more fundamental theory.

One of the most efficient ways to study top quarks will be to pair-produce them in future very energetic $e^+e^-$ colliders [1], through the reaction $e^+e^- \to t\bar{t}$. In general, the leptonic nature of the target allows for clean signals. In addition, due to their large masses, the produced top quarks are expected to decay weakly ($t\bar{t} \to bW^+\bar{b}W^-$, with subsequent leptonic decays of the $W$), before hadronization takes place; therefore electro-weak properties of the top can be studied in detail and QCD corrections can be reliably evaluated in the context of perturbation theory, when the energy of the collider is well above the threshold for $t\bar{t}$ production.

Motivated by this type of experiments, Atwood and Soni presented in a recent paper [2] a phenomenological analysis for determining the magnetic and electric dipole moment form factors of the top quark. Such form factors are defined through the following Lorentz decomposition of the $Vt\bar{t}$ vertex, where V represents a boson (a $\gamma$ or $Z$ in our case) coupled to the conserved leptonic current:

$$\Gamma_\mu^V(q^2) = \gamma_\mu F_1^V(q^2) + \frac{i\sigma_{\mu\nu}q^\nu}{2m_t}F_2^V(q^2) + \gamma_\mu\gamma_5 F_3^V(q^2) + \frac{i\sigma_{\mu\nu}q^\nu}{2m_t}\gamma_5 F_4^V(q^2) \tag{1.1}$$

where $\sigma_{\mu\nu} \equiv \frac{i}{2}[\gamma_\mu, \gamma_\nu]$ and $q^2 = s$, i.e. the Mandelstam variable associated to the squared energy of the center of mass. In the above decomposition, $F_2^V$ is the magnetic dipole moment (MDM) and $F_4^V$ is the electric dipole moment (EDM) form factor. In particular, $F_2^\gamma$ defined at $q^2 = 0$ is the usual definition of the anomalous magnetic moment. In the case of the top quark production, clearly $q^2 \geq 4m_t^2$. Within the SM the tree-level value for both $F_2^V$ and $F_4^V$ is zero. $F_2^V$ becomes non-zero through one-loop quantum corrections, whereas $F_4^V$, which violates $CP$, receives its first non-vanishing contributions at three loops [3]. The upshot of the analysis of [2] was that the dependence of the differential cross section for the reaction $e^+e^- \to t\bar{t}$ on the real and imaginary (absorptive) parts of the MDM and EDM form factors, for an incoming photon or $Z$, can be singled out *individually*, through a set of optimally chosen physical observables. Clearly, the possibility of such a detailed experimental study of the top dynamics is attractive, and could provide valuable probes for new physics.

The theoretical aspects of the situation are however not entirely clear. As it was pointed out already in the classic paper by Fujikawa, Lee, and Sanda [4], off-shell form factors of fermions are in general gauge dependent quantities. In the context of the $R_\xi$ gauges, for example, a residual dependence on the gauge-fixing parameter $\xi$ survives in the final expressions of form factors, when $q^2 \neq 0$ [5]. Obviously, in the case of $e^+e^-$ annihilation into heavy fermions, the value of $q^2$ must be above the heavy fermion threshold ($q^2 \geq 4m_t^2$, in our case). Consequently, the intermediate photon and $Z$ are far off-shell, and therefore, MDM and EDM form factors may in general be gauge-dependent and not suitable for comparison with experiment.

A way out of this problem is to define gauge independent off-shell vertices using the pinch technique (PT). The PT was invented by Cornwall over a decade ago [6] and has since been applied to a variety of physical problems. The main idea of this method is to resum via a well-defined algorithm the Feynman diagrams contributing to an ostensibly gauge-invariant process (like an S-matrix element), in such a way as to form new gauge-independent proper vertices, and new propagators with gauge-independent self-energies and only a trivial gauge dependence - that of their tree level parts. In the context of QCD a gauge invariant gluon self-energy was derived, and its Schwinger-Dyson equation constructed and solved for $T=0$ [7], as well as finite $T$ [8]. The plasmon decay rate was also calculated at finite $T$ using the same method [9]. Later the QCD gauge invariant three-gluon vertex was calculated at one loop level and was shown to satisfy a simple QED-like Ward identity [10]. The subleading corrections to the self-energy were calculated by Lavelle [11]. Finally, the gauge-invariant four-gluon QCD vertex was constructed and its Ward identity derived [12]. The PT was first extended to the case of non-Abelian gauge theories with spontaneously broken gauge symmetry (with elementary Higgs) in the context of a toy field theory based on $SU(2)$, and a gauge independent electro-magnetic form factor for the neutrino was constructed [13]. The complicated task of applying the P.T. in the electro-weak sector of the Standard Model was accomplished by Degrassi and Sirlin [14]. These last authors, in addition to deriving explicit expressions for the one loop gauge-invariant $WW$ and $ZZ$ self-energies, introduced an alternative description of the PT in terms of equal time commutators of currents. More recently [15], gauge independent $\gamma W^+W^-$ and $ZW^+W^-$ vertices were constructed in the context of the SM; they are related to the gauge-independent W-propagators introduced in [14] by a very simple Ward identity, and give rise to expressions for the W magnetic dipole and electric quadrupole moments, which, unlike previous treatments [16], are independent of the gauge-fixing parameter and infrared finite.

In this paper we focus on the MDM form factor of the top quark. In particular:

a. In the context of the SM we show that the conventionally defined form factor for the top quark is gauge dependent in the class of $R_\xi$ gauges. The gauge dependence is formally expressed as a double integral over the usual Feynman parameters, and is proportional to $\frac{q^2m_t^2}{M_w^4}$.

b. We show how one can define a *gauge independent* MDM form factor using the PT.

c. The gauge dependence found in (a) is computed and turns out to be numerically very strong; its presence distorts not only the quantitative but also the qualitative behavior of the answer obtained in (b). More specifically, unphysical thresholds are introduced, the asymptotic behavior of the form factors as $q^2 \to \infty$ is altered, and the numerical dominance of perturbative QCD, which is present in the gauge independent expressions of (b), is totally washed out. Of particular interest is the fact that the popular "unitary" gauge ( the limit of the $R_\xi$ gauges as $\xi \to \infty$) gives a completely wrong answer. This analysis indicates in retrospect that the gauge dependence established in (a) is a serious pathology and may lead to erroneous conclusions.

d. We extract the real and imaginary parts of the gauge independent MDM form factors for incoming photon and $Z$ ($F_2^\gamma$ and $F_2^Z$, respectively), and evaluate them numerically. We perform our computations for values of $q^2$ comfortably above the threshold for top production, and we ignore therefore non-perturbative effects due to threshold dynamics [17]. It turns out that even though perturbative QCD provides in general the dominant contribution, the MDM form factor displays a high sensitivity to the Higgs mass. It is important to emphasize that the Higgs dependence of the MDM is actually *stronger* than that displayed by the $e^+e^- \to t\bar{t}$ cross section as a whole [18].

The paper is organized as follows. In section 2 we review the S-matrix PT and discuss some of the more important results for our purposes. In addition, we briefly present Degrassi's and Sirlin's alternative formulation of the PT. In section 3 we perform an explicit calculation and determine the naive, gauge dependent expression for the MDM form factor. In section 4 we show how to use the pinch technique to define gauge independent $\gamma t\bar{t}$ and $Zt\bar{t}$ vertices, which give rise to gauge independent expressions for the MDM form factors. In section 5 we present a simple QED Ward identity relating the gauge independent $\gamma t\bar{t}$ vertex with the inverse top quark propagators, calculated in the Feynman gauge ($\xi = 1$). Section 6 contains the results of our gauge invariant MDM computations, and a discussion of their significance. We present our conclusions in section 7. Finally, in an Appendix, we list in detail the contributions of individual Feynman diagrams to the MDM form factors.

## 2. The Pinch Technique

In this section we briefly review the S-matrix pinch technique. In particular, we outline the method of derivation of the gauge-independent proper self-energy of a gauge boson and comment on the technical differences that arise when PT is applied to a theory with symmetry breaking, like the electro-weak theory, as opposed to a theory like QCD. In addition, we present the main idea of Degrassi's and Sirlin's formulation of the pinch, and establish some of the notation we will use in the sequel.

The PT is an algorithm that allows the construction of modified gauge independent n-point functions, through the order by order resummation of Feynman graphs contributing to a certain physical and therefore ostensibly gauge independent process (an S-matrix in our case). The simplest example that demonstrates how the P.T. works is the gauge boson two point function (propagator). Consider the $S$-matrix element $T$ for the elastic scattering of two fermions of masses $M_1$ and $M_2$. To any order in perturbation theory $T$ is independent of the gauge fixing parameter $\xi$, defined by the free gluon propagator

$$\Delta_{\mu\nu}(q) = \frac{-g_{\mu\nu} + (1-\xi)\frac{q_\mu q_\nu}{q^2}}{q^2} \tag{2.1}$$

On the other hand, as an explicit calculation shows, the conventionally defined proper self-energy (collectively depicted in graph 1a) depends on $\xi$. At the one loop level this dependence is cancelled by contributions from other graphs, like 1b and 1c, which do not seem to be of propagator type at first glance. That this must be so is evident from the form of $T$:

$$T(s,t,M_1,M_2) = T_1(t) + T_2(t,M_1,M_2) + T_3(s,t,M_1,M_2) \tag{2.2}$$

where the function $T_1(t)$ depends only on the Mandelstam variable $t = -(\hat{p}_1 - p_1)^2 = -q^2$, and not on $s = (p_1+p_2)^2$ or on the external masses. The propagator-like parts of graphs like 1e and 1f, which enforce the gauge independence of $T_1(t)$, are called "pinch parts". The pinch parts emerge every time a gluon propagator or an elementary three-gluon vertex contribute a longitudinal $k_\mu$ to the graph's numerator. The action of such a term is to trigger an elementary Ward identity of the form

$$\begin{aligned} k^\mu \gamma_\mu \equiv & \not{k} = (\not{p} + \not{k} - m) - (\not{p} - m) \\ = & S^{-1}(p+k) - S^{-1}(p) \end{aligned} \tag{2.3}$$

once it gets contracted with a $\gamma$ matrix. The first term on the right-hand side of Eq. (2.3) will remove the internal fermion propagator - that is a "pinch" - whereas $S^{-1}(p)$ vanishes on shell. This last property characterizes the S-matrix PT we will use throughout this paper. Returning to the decomposition of Eq. (2.2), the function $T_1(t)$ is gauge invariant and unique, and represents the contribution of the new propagator. We can construct the new propagator, or equivalently $T_1(t)$, directly from the Feynman rules. In doing so it is evident that any value for the gauge parameter $\xi$ may be chosen, since $T_1$, $T_2$, and $T_3$ are all independent of $\xi$. The simplest of all covariant gauges is certainly the Feynman gauge ($\xi = 1$), which gets rid of the longitudinal part of the gluon propagator. Therefore, the only possibility for pinching in this gauge arises from the four-momentum of the three-gluon vertices, and the only propagator-like contributions come from graph 1b.

To explicitly calculate the pinching contribution of a graph such as 1b it is convenient to decompose the vertex in the following way, first proposed by 't Hooft. Group theory factors aside,

$$\Gamma_{\mu\nu\alpha} = \Gamma^P_{\mu\nu\alpha} + \Gamma^F_{\mu\nu\alpha} \tag{2.4}$$

with

$$\Gamma^P_{\mu\nu\alpha} \equiv (q+k)_\nu g_{\mu\alpha} + k_\mu g_{\nu\alpha} \tag{2.5}$$

and

$$\Gamma^F_{\mu\nu\alpha} \equiv 2q_\mu g_{\nu\alpha} - 2q_\nu g_{\mu\alpha} - (2k+q)_\alpha g_{\mu\nu} \tag{2.6}$$

$\Gamma^F_{\mu\nu\alpha}$ satisfies a Feynman-gauge Ward identity:

$$q^\alpha \Gamma^F_{\mu\nu\alpha} = [k^2 - (k+q)^2] g_{\mu\nu} \tag{2.7}$$

where the RHS is the difference of two inverse propagators in the Feynman gauge. As for $\Gamma^{P}_{\mu\nu\alpha}$ (P for "pinch"), it gives rise to pinch parts when contracted with $\gamma$ matrices

$$\begin{aligned} g_{\mu\alpha}(\not{q}+\not{k}) =& g_{\mu\alpha}[(\not{p}+\not{q}-m)-(\not{p}-\not{k}-m)] \\ =& g_{\mu\alpha}[S^{-1}(p+q)-S^{-1}(p-k)] \end{aligned} \tag{2.8}$$

and

$$\begin{aligned} g_{\nu\alpha}\not{k} =& g_{\nu\alpha}[(\not{p}-m)-(\not{p}-\not{k}-m)] \\ =& g_{\nu\alpha}[S^{-1}(p)-S^{-1}(p-k)] \end{aligned} \tag{2.9}$$

Now both $S^{-1}(p+q)$ and $S^{-1}(p)$ vanish on shell, whereas the two terms proportional to $S^{-1}(p-k)$ pinch out the internal fermion propagator in graph 1b, and so we are left with two "pinch" (propagator-like) parts and one "regular" (purely vertex-like) part:

$$pinch\ \ part = 2\times(-\frac{1}{2}c_A)[\frac{ig^2}{(2\pi)^4}]\int\frac{d^4k}{k^2(k+q)^2}(g^{\mu\rho}\gamma_\rho) \tag{2.10}$$

$$regular\ \ part = (\frac{1}{2}c_A)[\frac{ig^2}{(2\pi)^4}]\int\frac{d^4k\gamma^\rho S(p-k)\gamma^\sigma\Gamma^F_{\rho\sigma\mu}}{k^2(k+q)^2} \tag{2.11}$$

with $c_A$ the Casimir operator for the adjoint representation [ $c_A = N$ in $SU(N)$ ] and one factor of 2 for the two pinching terms from Eq. (2.8) and Eq. (2.9). When we add to the usual propagator graphs (in the Feynman gauge) the pinch contributions of Eq. (2.10), together with an equal contribution coming from the mirror-image graph of 1.b, we find the new gauge invariant self-energy $\hat{\Pi}_{\mu\nu}(q)$, given by

$$\hat{\Pi}_{\mu\nu}(q) = (q^2g_{\mu\nu}-q_\mu q_\nu)\hat{\Pi}(q) \tag{2.12}$$

where

$$\hat{\Pi}(q) = -bg^2\ln(\frac{-q^2}{\mu^2}) \tag{2.13}$$

and $b=\frac{11N}{48\pi^2}$ is the coefficient in front of $-g^3$ in the usual one loop $\beta$-function. We see that the modified propagator has a gauge independent self-energy and only a trivial gauge dependence originating from the tree part given by Eq. (2.1). After all pinch contributions from graph 1.b have been allotted to the new gluon self-energy, the rest of graph 1.b, namely the expression in Eq. (2.11), is genuinely vertex-like and must be added to the usual QED-like graphs (Fig.2a). The final gauge-invariant vertex $\hat{\Gamma}^a_\alpha$ is given by

$$\hat{\Gamma}^a_\mu = \frac{i\tau_\alpha g^2}{(2\pi)^4}[(\frac{c_A}{2})\int \frac{d^4k\gamma^\rho S(p-k)\gamma^\sigma\Gamma^F_{\rho\sigma\mu}}{k^2(k+q)^2}+(\frac{c_A}{2}+\frac{Nd_f}{c_f})\int \frac{d^4k\gamma^\rho S(p+q-k)\gamma_\mu S(p-k)\gamma_\rho}{k^2}] \tag{2.14}$$

where $\tau_\alpha$ is the fermion representation matrix, $d_f$ its dimension, and $C_f$ its Dynkin index. If we now act on the $\hat{\Gamma}^a_\alpha$ given in Eq. (2.14) with $q^\alpha$ and use Eq. (2.7) it immediately follows that $\hat{\Gamma}^a_\alpha$ satisfies the following QED-like Ward identity:

$$q^\mu\hat{\Gamma}^a_\mu = \tau^a[\Sigma(p) - \Sigma(\hat{p})] \tag{2.15}$$

with $\Sigma$ the usual quark self-energy in the Feynman gauge and $\hat{p} = p + q$. [19] The application of the P.T. in a theory with spontaneous symmetry breaking like the Standard Model is significantly more involved [13,14]. The main reason is that the massive charged vector mesons couple to fermions with different masses, and therefore the elementary Ward identity of Eq. (2.3) gets modified to:

$$\begin{aligned} \not{k} &= (\not{p}+\not{k}-m_1) - (\not{p}-m_2) + (m_1-m_2) \\ &= S^{-1}(p+k) - S^{-1}(p) + (m_1-m_2) \end{aligned} \tag{2.16}$$

The first two terms of Eq. (2.16) will pinch and vanish on shell, respectively, as they did before. But in addition a term proportional to $m_1 - m_2$ is left over. However, as it was shown in [13], any additional gauge-dependent contributions coming from Eq. (2.16) cancel exactly against contributions coming from graphs involving charged Goldstone bosons, whose couplings to fermions are also proportional to $m_1 - m_2$.

Finally, we conclude this section with a brief presentation of an alternative formulation of the PT introduced in [14] in the context of the Standard Model. In this approach the interaction of gauge bosons with external fermions is described in terms of current correlation functions, i.e. matrix elements of Fourier transforms of time-ordered products of current operators [20]. This is particularly economical because these amplitudes automatically include several closely related Feynman diagrams. When one of the current operators is contracted with the appropriate four-momentum, a Ward identity is triggered. The pinch part is then identified with the contributions involving the equal-time commutators in the Ward identities, and therefore involve amplitudes in which the number of current operators has been decreased by one or more. A basic ingredient in this formulation are the following equal-time commutators, some of which we will also employ later in section 3:

$$\delta(x_0-y_0)[J_W^0(x),J_Z^\mu(y)]=c^2J_W^\mu(x)\delta^4(x-y) \tag{2.17}$$

$$\delta(x_0-y_0)[J_W^0(x),J_W^{\mu\dagger}(y)]=-J_3^\mu(x)\delta^4(x-y) \tag{2.18}$$

$$\delta(x_0-y_0)[J_W^0(x),J_\gamma^\mu(y)]=J_W^\mu(x)\delta^4(x-y) \tag{2.19}$$

with $J_3^\mu\equiv 2(J_Z^\mu+s^2J_\gamma^\mu)$. On the other hand

$$\delta(x_0-y_0)[J_{V_i}^0(x),J_{V_j}^\mu(y)]=0 \tag{2.20}$$

where $V_i,V_j\in\{\gamma,Z\}$ To demonstrate the method with an example, consider the vertex $\Gamma_\mu$ shown in Fig.1b, where now the gauge particles in the loop are W s instead of gluons and the incoming and outgoing fermions are massless. It can be written as follows (with $\xi=1$):

$$\Gamma_\mu=\int\frac{d^4k}{2\pi^4}\Gamma_{\mu\alpha\beta}(q,k,-k-q)\int d^4xe^{ikx}<f|T^*[J_W^{\alpha\dagger}(x)J_W^\beta(0)]|i> \tag{2.21}$$

When an appropriate momentum, say $k_\alpha$, from the vertex is pushed into the integral over $dx$, it gets transformed into a covariant derivative $\frac{d}{dx_\alpha}$ acting on the time ordered product $<f|T^*[J_W^{\alpha\dagger}(x)J_W^\beta(0)]|i>$. After using current conservation and differentiating with respect to the $\theta$-function terms, implicit in the definition of the $T^*$ product, we end up with the left-hand side of Eq. (2.18). So, the contribution of each such term is proportional to the matrix element of a single current operator, namely $< f|J_3^{\mu}|i >$; this is precisely the pinch part. Calling $\Gamma_{\mu}^{P}$ the total pinch contribution from the $\Gamma_{\mu}$ of Eq. (2.21), we find that

$$\Gamma_{\mu}^{P} = -g^3 c I_{WW}(Q^2) < f|J_3^{\mu}|i > \tag{2.22}$$

where

$$I_{ij}(q) = i\int (\frac{d^4k}{2\pi^4}) \frac{1}{(k^2 - M_i^2)[(k+q)^2 - M_j^2]} \tag{2.23}$$

Obviously, the integral in Eq. (2.23) is the generalization of the QCD expression Eq. (2.8) to the case of massive gauge bosons.

## 3. Gauge dependence of the conventional MDM form factor.

In this section we explicitly show that the MDM form factor extracted from the conventional vertex graphs, calculated in a general renormalizable ($R_{\xi}$) gauge, depends on the gauge-fixing parameter $\xi$, and is therefore not suitable for comparison with experiment. We then isolate the gauge dependent contribution to the MDM originating from the vertex-like pinch parts of box graphs, calculated in the same class of gauges. It turns out that the latter gauge-dependent contributions to the MDM cancel *exactly* against the gauge-dependent parts of the conventional vertex graphs, thus giving rise to a gauge independent expression for the top MDM. It is important to emphasize that even though it has been proved [14] that the inclusion of the vertex-like pinch parts is necessary in order to define a gauge independent $\gamma t\bar{t}$ vertex, it has not been shown explicitly that the conventionally defined expression for the MDM is indeed *gauge dependent*. The obvious alternative (which we will show not to be true) is that, even though the usually defined vertex has an overall gauge dependence, its part proportional to $\sigma_{\mu\nu}q^{\nu}$ could be fortuitously gauge independent. In addition, the explicit computation of the gauge dependence of the conventionally defined

MDM enables one to appreciate quantitatively the significance of the issue at hand. The tree-level vector-meson propagator $\Delta_{\mu\nu}(k)^i$ in a general covariant gauge is given by

$$\Delta^i_{\mu\nu}(k) = [g_{\mu\nu} - (1-\xi_i)\frac{k_\mu k_\nu}{k^2 - \xi_i M_i^2}]\frac{1}{k^2 - M_i^2} \tag{3.1}$$

with $i = w, z, \gamma$, and $M_\gamma = 0$. The propagator $D_c(k)$ of the two charged Goldstone bosons is given by

$$D_c(k) = \frac{1}{k^2 - \xi_w M_w^2} \tag{3.2}$$

and the propagator $D_n(k)$ of the neutral Goldstone boson is

$$D_n(k) = \frac{1}{k^2 - \xi_z M_z^2} \tag{3.3}$$

The propagator $D_H(k)$ of the physical Higgs particle is gauge-independent at tree-level

$$D_H(k) = \frac{1}{k^2 - m_H^2} \tag{3.4}$$

and therefore the graph of Fig.3.16 has no gauge-dependent contribution. Finally, we denote the quark propagators by

$$S_j(k) = \frac{1}{\not{k} - m_j} \tag{3.5}$$

with $j = t, b$, for top and bottom quarks. We further define $c = cos\theta_W$ and $s = sin\theta_W$, where $\theta_W$ is the Weinberg mixing angle, $\kappa = |U_{tb}|^2$ for the modulus of the C.K.M. mixing between top and bottom [21], and $C_Z = 1 - \frac{8}{3}s^2 - \gamma_5$. Finally, the left-right projectors are given by $P_{L,R} = 1 \mp \gamma_5$

We denote by $V_\mu^{(i)}$ the $\xi$-dependent parts of the graphs $\Gamma_\mu^{(i)}$, shown in Fig.3, with $i = 1, 2, .., 16$

We use the identity

$$\frac{(\xi - 1)}{(k^2 - M^2)(k^2 - \xi M^2)} = \frac{1}{M^2}[\frac{1}{k^2 - \xi M^2} - \frac{1}{k^2 - M^2}] \tag{3.6}$$

in order to isolate the $\xi$-dependent parts of the graphs. Gauge-dependent terms proportional to $q^{\mu}$ are set to zero, when considered "sandwiched" between the on-shell target electrons. In what follows we omit the loop momentum integration symbol $(\frac{e}{2s\sqrt{2}})^2\int\frac{d^4k}{2\pi^4}$, and we use dimensional regularization whenever necessary. We define the scalar quantities $D_1(\xi_i)$, $D_2(\xi_i)$, and $D_3(\xi_i)$ as follows

$$
\begin{aligned}
D_1(\xi_i) &= \frac{1}{k^2-\xi_i M_i^2} \\
D_2(\xi_i) &= \frac{1}{[(k+q)^2-M_i^2](k^2-\xi_i M_i^2)} \\
D_3(\xi_i) &= \frac{1}{[(k+q)^2-\xi_i M_i^2](k^2-\xi_i M_i^2)}
\end{aligned}
\tag{3.7}
$$

Clearly, $D_2(\xi_i=1)=D_3(\xi_i=1)$.

The most involved graph is $\Gamma_{\mu}^{(1)}$ of Fig.3, given by

$$
\Gamma_{\mu}^{(1)} = \kappa\Gamma_{\mu\tau\lambda}\Delta^{\tau\rho}(k+q)\Delta^{\lambda\sigma}(k)\gamma_{\rho}P_L S_b(k+p_2)\gamma_{\sigma}P_L \tag{3.8}
$$

Its $\xi_w$-dependent part $V_{\mu}^{P}(\xi_w)$ is given by

$$
V_{\mu}^{P}(\xi_w) = q^2[V_{\mu}^{1P}+V_{\mu}^{2P}+V_{\mu}^{3P}] \tag{3.9}
$$

with

$$
\begin{aligned}
V_{\mu}^{1P} &= \kappa(1-\xi_w)\frac{\gamma_{\mu}P_L S_b(k+p_2)[m_t P_R - m_b P_L]}{[(k+q)^2-M_w^2][(k+q)^2-\xi_w M_w^2](k^2-M_w^2)} \\
V_{\mu}^{2P} &= \kappa(1-\xi_w)\frac{[m_b P_R - m_t P_L]S_b(k+p_2)\gamma_{\mu}P_L}{[(k+q)^2-M_w^2](k^2-M_w^2)(k^2-\xi_w M_w^2)} \\
V_{\mu}^{3P} &= \kappa(1-\xi_w)^2\frac{k_{\mu}[m_b P_R - m_t P_L]S_b(k+p_2)[m_t P_R - m_b P_L]}{[(k+q)^2-M_w^2][(k+q)^2-\xi_w M_w^2](k^2-M_w^2)(k^2-\xi_w M_w^2)}
\end{aligned}
\tag{3.10}
$$

The superscript $P$ in the formulas above stands for "pinch"; it hints to the fact that these contributions will eventually cancel against pinch parts coming from box graphs (see next section). It is important to notice the presence of the factor $q^2$ in the r.h.s. of Eq. (3.9). Clearly, $V_{\mu}^{P}(\xi_w=1)=0$.

The $\xi$-dependent contributions $V_\mu^{(i)}$ from each individual graph $\Gamma_\mu^{(i)}$ are:

$$
\begin{aligned}
V_\mu^{(1)} &= V_\mu^P(\xi_w) - V_\mu^{(2)} - V_\mu^{(3)} \\
V_\mu^{(2)} &= \frac{\kappa}{M_w}\gamma_\mu P_L S_b(k+p_2)[m_t P_R - m_b P_L] D_2(\xi_w) - \frac{1}{2}V_\mu^{(4)} \\
V_\mu^{(3)} &= \frac{\kappa}{M_w}[m_b P_R - m_t P_L] S_b(k+p_2)\gamma_\mu P_L D_2(\xi_w) - \frac{1}{2}V_\mu^{(4)} \\
V_\mu^{(4)} &= \frac{2\kappa}{M_w^2} k_\mu [m_b P_R - m_t P_L] S_b(k+p_2)[m_t P_R - m_b P_L] D_3(\xi_w)
\end{aligned}
$$

$$
\begin{aligned}
V_\mu^{(5)} =& \frac{\kappa}{3M_w^2}\Bigg[\gamma_\mu P_L S_b(k+p_1)[m_t P_R - m_b P_L] - [m_b P_R - m_t P_L] S_b(k+p_2)\gamma_\mu P_L \\
&+ 2\gamma_\mu P_L D_1(\xi_w)\Bigg] - V_\mu^{(8)}
\end{aligned}
\tag{3.11}
$$

$$
\begin{aligned}
V_\mu^{(6)} &= \frac{1}{3M_w^2}[\gamma_\mu C_Z^2 - 2m_t\gamma_\mu C_Z S_t(k+p_1)\gamma_5 + 2m_t\gamma_5 S_t(k+p_2)\gamma_\mu C_Z] D_1(\xi_z) - V_\mu^{(9)} \\
V_\mu^{(7)} &= -U_\mu^{(3)} \\
V_\mu^{(8)} &= \frac{1}{3}(\frac{\kappa}{M_w^2})[m_b P_R - m_t P_L] S_t(k+p_2)\gamma_\mu S_t(k+p_1)[m_t P_R - m_b P_L] D_1(\xi_w) \\
V_\mu^{(9)} &= \frac{4}{3}(\frac{m_t^2}{M_w^2})\gamma_5 S_t(k+p_2)\gamma_\mu S_t(k+p_1)\gamma_5 D_1(\xi_z) \\
V_\mu^{(10)} &= 0
\end{aligned}
$$

The analogous $\xi$-dependent contributions from the self-energy graphs $\Sigma_t^{(i)}$ of the external top quarks (Fig.(3)) are:

$$
\begin{aligned}
V_\mu^{(11)} \;\; = \;\; & -\frac{\kappa}{3M_w^2}\Bigg[\gamma_\mu P_L S_b(k+p_1)[m_t P_R - m_b P_L] - [m_b P_R - m_t P_L] S_b(k+p_2)\gamma_\mu P_L \\
&+ 2\gamma_\mu P_L D_1(\xi_w)\Bigg] - V_\mu^{(14)}
\end{aligned}
$$

$$
\begin{aligned}
V_\mu^{(12)} &= -\frac{1}{3M_w^2}[\gamma_\mu C_Z^2 - 2m_t\gamma_\mu C_Z S_t(k+p_1)\gamma_5 + 2m_t\gamma_5 S_t(k+p_2)\gamma_\mu C_Z] D_1(\xi_z) - V_\mu^{(15)} \\
V_\mu^{(13)} &= -(1-\xi_\gamma)(\frac{4}{3})^3\gamma_\mu\frac{1}{k^4}
\end{aligned}
$$

$$V_\mu^{(14)} = \frac{1}{3}(\frac{\kappa}{M_w^2})\Bigg[\gamma_\mu S_t(p_1)[m_b P_R - m_t P_L]S_t(k+p_1)[m_t P_R - m_b P_L] + \\ + [m_b P_R - m_t P_L]S_t(k+p_2)[m_t P_R - m_b P_L]S_t(p_2)\gamma_\mu\Bigg] D_1(\xi_w) \tag{3.12}$$

$$V_\mu^{(15)} = \frac{4}{3}(\frac{m_t^2}{M_w^2})[\gamma_\mu S_t(p_1)\gamma_5 S_t(k+p_1)\gamma_5 + \gamma_5 S_t(k+p_2)\gamma_5 S_t(p_2)\gamma_\mu]D_1(\xi_z)$$

$$V_\mu^{(16)} = 0$$

Adding the above equations together we see that $\sum_{i=1}^{16} V_\mu^{(i)} = V_\mu^P$, so that

$$\Gamma_\mu^{(\xi_w)} = \Gamma_\mu^R|_{(\xi_w=1)} + V_\mu^P(\xi_w) \tag{3.13}$$

In the equation above $\Gamma_\mu^R|_{(\xi_w=1)}$ represents the "regular"(hence the superscript R), purely vertex-like contributions in the Feynman gauge ($\xi_w = 1$). The term "regular" or "purely vertex-like" refers to what is left from the vertex graphs, after the propagator-like pieces have been removed through pinching. (see Eq. (2.11) and Eq. (2.14))

The calculation of the gauge dependence of the $Zt\bar{t}$ vertex $\Lambda_\mu$ proceeds in an analogous way. The final answer is exactly the same as in Eq. (3.13), except that one must replace $V_\mu^P$ by $V_\mu^{Pz}$ defined as

$$V_\mu^{Pz} = (q^2 - M_z^2)[V_\mu^{1P} + V_\mu^{2P} + V_\mu^{3P}], \tag{3.14}$$

e.g.

$$\Lambda_\mu^{(\xi_w)} = \Lambda_\mu^R|_{(\xi_w=1)} + V_\mu^{Pz}(\xi_w) \tag{3.15}$$

It is important to notice the particular form of the gauge-dependent terms $V_\mu^P(\xi_w)$ and $V_\mu^{Pz}$, namely that they are explicitly multiplied by $q^2$ and $q^2 - M_z^2$, respectively. As we will see in the next section, this is precisely the characteristic structure of vertex-like contributions originating from box graphs, after pinching. Moreover, the gauge-dependent parts vanish, when the incoming gauge bosons are on-shell ($q^2 = 0$ and $q^2 = M_z^2$, respectively), as they should.

We next proceed to extract from Eq. (3.13) the form factor $F(q^2)$ proportional to $\sigma_{\mu\nu}q^\nu$. We use the Gordon decomposition

$$\bar{u}(p_2)\gamma_\mu u(p_1) = \frac{1}{2m_t}\bar{u}(p_2)[(p_1+p_2)_\mu + \frac{i\sigma_{\mu\nu}q^\nu}{2m_t}]u(p_1) \tag{3.16}$$

to convert terms proportional to $(p_1+p_2)_\mu$ into linear combinations proportional to $\gamma_\mu$ and $\sigma_{\mu\nu}q^\nu$. The final answer is:

$$F_2(q^2) = F_2(q^2)|_{(\xi_w=1)} + f_2(\xi_w, q^2) \tag{3.17}$$

where $F_2(q^2)|_{(\xi_w=1)}$ originates from the first term in the r.h.s. of Eq. (3.13), whereas $f_2(\xi_w, q^2)$ originates from the second one, and has the following closed form:

$$f_2(\xi_w, q^2) = \frac{2q^2m_t^2}{M_w^2}\int_0^1 dx\int_0^{1-x} dy\Bigg[xR(1,\xi_w) + yR(\xi_w,1) - (x+y)R(1,1) + [\frac{m_b^2}{M_w^2}(x+y) - \frac{m_t^2+m_b^2}{2M_w^2}(x+y)^2][R(1,1)+R(\xi_w,\xi_w)-R(1,\xi_w)-R(\xi_w,1)]\Bigg] \tag{3.18}$$

with

$$R(\xi_1,\xi_2) = [(x+y)^2m_t^2 - q^2xy - (x+y)(m_t^2+m_b^2) + x\xi_1M_w^2 + y\xi_2M_w^2 + m_b^2]^{-1} \tag{3.19}$$

Clearly, $F_2(q^2)$ of Eq. (3.17) is $\xi_w$-dependent, through the explicit dependence of $f_2(\xi_w, q^2)$ on $\xi_w$. This concludes the proof that the conventionally defined MDM form factor is gauge-dependent, for both an incoming photon or $Z$, and therefore unsuitable in general for comparison with experiment.

It is interesting to mention that there is *no* contribution from either parts of Eq. (3.13) proportional to the CP violating term $\sigma_{\mu\nu}q^\nu\gamma_5$. Therefore, the usual statement that there is no one-loop contribution to the EDM is indeed a gauge-independent one.

## 4. Gauge-independent MDM form factors via the PT

In this section we show how the use of the PT can give rise to gauge-independent expressions for the MDM form factors, for both an incoming photon and $Z$. The important point is to recognize that the box-like graphs, like the one shown in Fig.2b, contain gauge-dependent contributions which are kinematically equivalent to a $\gamma t\bar{t}$ or a $Zt\bar{t}$ vertex. When all these contributions are added to the conventional vertex graphs, the gauge dependence cancels exactly.

The only kinematically relevant box graph is the one with an intermediate $W^+W^-$ pair, shown in Fig.2b . [22]. Its $\xi_w$-dependent contributions $B$ are given by [23]

$$B(\xi_w) = g^2 e^- \gamma_\nu P_L e^+ g^{\nu\mu} [V_\mu^{1P} + V_\mu^{2P} + V_\mu^{3P}] + ... \tag{4.1}$$

The dots in Eq. (4.1) indicate terms that belong to one of the following categories:

a. Propagator-like, to be allotted to the $\gamma\gamma$, $\gamma Z$ $Z\gamma$, and $ZZ$ self-energies.

b. Vertex-like of the form $g^2[V_\nu^{1P} + V_\nu^{2P} + V_\nu^{3P}] g^{\nu\mu} \bar{t}\gamma_\mu P_L t$; the $V_\nu^{1P}$, $V_\nu^{2P}$ and $V_\nu^{3P}$ are given in Eq. (3.10) after replacing $t \leftrightarrow e$ and $b \leftrightarrow \nu$. These contribution can be combined with the conventional $\gamma e^+e^-$ and $Ze^+e^-$ vertex graphs to construct gauge-independent $\gamma e^+e^-$ and $Ze^+e^-$ vertices.

c. Purely box-like, to be combined with the rest of the box graphs.

Clearly, none of the categories listed above contributes kinematically to the MDM form factors.

We now write the expression given in Eq. (4.1) as a linear combination of a $\gamma e^+e^-$ and a $Ze^+e^-$ bare vertex, using the simple identity

$$\frac{g^2}{4} e^+ \gamma_\mu P_L e^- = g^2 s^2 e^+ \gamma_\mu e^- + \frac{g^2}{4} e^+ \gamma_\mu [P_L - 4s^2] e^- \tag{4.2}$$

so that $B$ becomes (we omit the external $e^+$ and $e^-$ spinors)

$$\begin{aligned} B = & -\frac{g^2 s^2 \gamma^\mu}{q^2}[q^2(V_\mu^{1P}+V_\mu^{2P}+V_\mu^{3P})] - \frac{(\frac{g^2}{4})\gamma^\mu[P_L - 4s^2]}{q^2-M_z^2}[(q^2-M_z^2)(V_\mu^{1P}+V_\mu^{2P}+V_\mu^{3P})] \\ = & -\frac{g^2 s^2 \gamma^\mu}{q^2}V_\mu^{P} - \frac{(\frac{g^2}{4})\gamma^\mu[P_L-4s^2]}{q^2-M_z^2}V_\mu^{Pz} \end{aligned} \tag{4.3}$$

If we now add the first term on the r.h.s. of Eq. (4.3) to Eq. (3.13), and the second term to Eq. (3.15), we find the gauge-independent $\gamma t\bar{t}$ and $Zt\bar{t}$ vertices, $\hat{\Gamma}_\mu$ and $\hat{\Lambda}_\mu$, respectively. They are simply

$$\hat{\Gamma}_\mu = \Gamma_\mu^R|_{(\xi_w=1)} \tag{4.4}$$

and

$$\hat{\Lambda}_\mu = \Lambda_\mu^R|_{(\xi_w=1)} \tag{4.5}$$

thus, the gauge-independent MDM form factor $\hat{F}(q^2)$ is given by:

$$\hat{F}(q^2) = F(q^2)|_{\xi_w=1} \tag{4.6}$$

for both incoming photon and $Z$.

The *gauge-independent* $\gamma t\bar{t}$ and $Zt\bar{t}$ vertices given in Eq. (4.4) and Eq. (4.5) coincide with the the conventional vertex graphs of Fig.3, calculated in the $\xi_w = 1$ gauge, but with the propagator-like pieces of graph 3.1 removed through pinching. In particular, the gauge-independent MDM form factors can be obtained if one just calculates the *conventional* MDM form factors in that same gauge. This results may come as no surprise to the reader familiar with the PT. Indeed, as we explained in section 2, in calculating gauge-independent Green's functions (a vertex in our case) any gauge choice is as good as any other, as long as one identifies and judiciously allots the pinch contributions. If one decides to calculate the gauge-independent vertex in the $\xi_w = 1$ gauge, the only possibility for pinching comes from the three-boson vertex of graph 3.1, which gives rise to a propagator-like piece proportional to $\gamma_\mu P_L$. In particular, there are no pinch contributions (propagator or vertex-like) coming from box graphs, since there are no longitudinal terms in the bare gauge boson propagators appearing in the boxes. Since the only term removed from the vertex has no part proportional to $\sigma_{\mu\nu}q^{\nu}$, and hence no contribution to the MDM, the answer for the gauge-independent MDM is obviously the one recorded in Eq. (4.6). The fact that the gauge-independent answer in the MDM case coincides with the result in the Feynman gauge is in a sense fortuitous, and is certainly not true in general. Indeed, several other gauge-independent modified Green's functions, constructed via the PT, turned out to be *different* from the corresponding result in the Feynman gauge, or any other popular gauge for that matter.

Finally, we conclude this section with a comparison of the gauge-independent ($\xi_w = 1$) answer for the MDM, to the answers computed for different values of $\xi_w$, as given in Eq. (3.17) and Eq. (3.18). We chose $m_t = 130\ GeV$ and $M_H = 150\ GeV$. The results shown in Fig.4 are rather impressive and convincingly demonstrate the importance of the issue at hand. Indeed, the residual gauge dependence is not just a theoretical nuisance, but has a huge impact on the predictions of the theory, affecting them both quantitatively and qualitatively. In Fig.3 the real part of the MDM form factor for an incoming photon is plotted as a function of $\sqrt{s}$. The curve marked (a) is the gauge-independent answer, (b) corresponds to the gauge choice $\xi_w = 10$, (c) to $\xi_w = 100$, (d) to $\xi_w = 1000$, and (e) to the unitary gauge ($\xi_w \to \infty$) [24]. We see how the curves start approaching the unitary limit (e), as $\xi_w$ becomes large. The gauge-dependent answers are typically at least one order of magnitude *larger* than the gauge-independent answer of (a). The reason for this artificial enhancement is the presence of the factor $\frac{q^2}{M_w^2}$ in front of the expressions in Eq. (3.18), which is huge due to the massiveness of the top quark. Indeed, already at the threshold of top pair production, for $m_t = 130\ GeV$, we have that $\frac{q^2}{M_w^2} = 4\frac{m_t^2}{M_w^2} \approx 10.5$. If instead of $t\bar{t}$ we considered $b\bar{b}$ production the corresponding factor would be of the order of $10^{-2}$. The phenomenological implications of such an enhancement are clear. For example, any dependence of the final answer on the mass of the Higgs boson gets totally altered.

Because of the same reason, the asymptotic behavior of the MDM form factor in a generic $R_\xi$ gauge is totally different than that of the gauge independent case; as we see from Fig.4, the MDM curves increase monotonically as $\sqrt{s} \to \infty$, whereas, on the contrary, the gauge-independent curve approaches zero. It is interesting to notice that within the entire class of $R_\xi$ gauges, only the gauge-independent answer decreases asymptotically [25]. In addition, as we can see in curves (b) and (c), extra $\xi$-dependent thresholds are introduced, which are artifacts and do not correspond to anything physical. Even though it is very unlikely that anyone would calculate in any of the finite gauges ($\xi = 10, 100, 1000$) considered above, it is important to realize that the popular unitary gauge (e) yields a totally wrong answer. The same is true for another popular gauge, the Landau gauge, with $\xi_w = 0$ (not shown). There, in addition to the usual artifacts, one has extra infrared divergences, due to the presence of massless poles.

## 5. The Ward identity for the gauge-independent $\gamma t\bar{t}$ vertex

In this section we prove a QED-like Ward identity, which relates the gauge-independent $\gamma t\bar{t}$ vertex $\hat{\Gamma}_\mu$ to the top quark self-energy $\Sigma_t$ in the Feynman gauge. The Ward identity is

$$q^\mu \hat{\Gamma}_\mu(p_1, p_2) = \frac{2}{3}[\Sigma_t(p_2) - \Sigma_t(p_1)] \tag{5.1}$$

which is the one-loop generalization of the trivial tree-level result $Q_t q^\mu \gamma_\mu = Q_t[(\not{p}_2 - m_t) - (\not{p}_1 - m_t)]$, with $Q_t = \frac{2}{3}$ the electric charge of the top quark. It is important to emphasize that it is only after the propagator-like contributions of graph (3.1) have been removed that the validity of Eq. (5.1) becomes possible. To see that, let us recall the decomposition of the tree-level $\gamma W^+ W^-$ vertex, given by Eq. (2.5) and Eq. (2.6). The graph $\Gamma_\mu^{(1)}$, in the Feynman gauge, is (we again omit the $(\frac{e}{2s\sqrt{2}})^2 \int \frac{d^4k}{2\pi^4}$ factor in front):

$$\Gamma_\mu^{(1)} = \kappa(\Gamma^P_{\mu\sigma\rho} + \Gamma^F_{\mu\sigma\rho})\gamma^\sigma P_L \frac{1}{\not{k} - m_b}\gamma^\rho P_L [\frac{1}{(k^2 - M_w^2)[(k+q)^2 - M_w^2]}] \tag{5.2}$$

The part of Eq. (5.2) proportional to $\Gamma^{P}_{\mu\sigma\rho}$ , after pinching out the internal quark propagator, gives rise to propagator-like contributions, which, as explained in section 2, will be allotted to the self-energies of the incoming neutral gauge bosons ($\gamma\gamma, ZZ, \gamma Z, Z\gamma$). However, unlike the QCD case, a residual contribution $\Gamma^{R}_{\mu}$ survives, because the internal ($b$) and external ($t$) quarks have different masses. This extra contribution is

$$\begin{aligned}\Gamma^{R}_{\mu} =& \frac{\{[m_t P_R - m_b P_L]S_b(k)\gamma_\mu P_L + \gamma_\mu P_L S_b(k)[m_b P_R - m_t P_L]\}}{(k^2 - M_w^2)[(k+q)^2 - M_w^2]} \\ =& -(\Gamma^{(2)}_{\mu} + \Gamma^{(3)}_{\mu})\end{aligned} \tag{5.3}$$

and cancels entirely against the contributions of graphs $\Gamma^{(2)}_{\mu}$ and $\Gamma^{(3)}_{\mu}$. The above observation is crucial for deriving Eq. (5.1). So, the purely vertex-like contribution $\Gamma^{(VL)}_{\mu}$ from $\Gamma^{(1)}_{\mu}$ is given by

$$\Gamma^{(VL)}_{\mu} = \kappa \frac{\Gamma^{F}_{\mu\sigma\rho}\gamma^\sigma P_L S_b(k)\gamma^\rho P_L}{(k^2 - M_w^2)[(k+q)^2 - M_w^2]} - (\Gamma^{(2)}_{\mu} + \Gamma^{(3)}_{\mu}) \tag{5.4}$$

After these considerations, proving the Ward identity is rather straightforward. In that vein, it is far more convenient to act with $q^\mu$ on each graph individually, instead of first calculating them, and then acting with $q^\mu$ on the final answer.Using Eq. (2.7) we have:

$$\begin{aligned}
q^\mu[\Gamma^{(VL)}_{\mu} + \Gamma^{(2)}_{\mu} + \Gamma^{(3)}_{\mu}] &= \Sigma^{(1)}_t(p_2) - \Sigma^{(1)}_t(p_1) \\
q^\mu\Gamma^{(4)}_{\mu} &= \Sigma^{(4)}_t(p_2) - \Sigma^{(4)}_t(p_1) \\
q^\mu\Gamma^{(5)}_{\mu} &= -\frac{1}{3}[\Sigma^{(1)}_t(p_2) - \Sigma^{(1)}_t(p_1)] \\
q^\mu\Gamma^{(6)}_{\mu} &= \frac{2}{3}[\Sigma^{(2)}_t(p_2) - \Sigma^{(2)}_t(p_1)] \\
q^\mu\Gamma^{(7)}_{\mu} &= \frac{2}{3}[\Sigma^{(3)}_t(p_2) - \Sigma^{(3)}_t(p_1)] \\
q^\mu\Gamma^{(8)}_{\mu} &= -\frac{1}{3}[\Sigma^{(4)}_t(p_2) - \Sigma^{(4)}_t(p_1)] \\
q^\mu\Gamma^{(9)}_{\mu} &= \frac{2}{3}[\Sigma^{(5)}_t(p_2) - \Sigma^{(5)}_t(p_1)] \\
q^\mu\Gamma^{(10)}_{\mu} &= \frac{2}{3}[\Sigma^{(6)}_t(p_2) - \Sigma^{(6)}_t(p_1)]
\end{aligned} \tag{5.5}$$

Adding the left and right hand sides of Eq. (5.5) we arrive at the advertised result. Clearly, pinching is instrumental for the validity of Eq. (5.1).

## 6. Calculations and results

In this section we report the results of the numerical evaluation of the integrals listed in the Appendix. These integrals correspond to the vertex graphs of Fig.3, calculated in the $\xi = 1$ gauge, which as we explained in section 4, *coincides* with the gauge-independent answer in the PT framework. In arriving at these expressions we used the on-shell conditions $\not{p}_1 = \not{p}_2 = m_t$ (or equivallently $p_1^2 = p_2^2 = m_t^2$), and the Gordon decomposition of Eq. (3.16).

We have used very accurate integration routines contained in the Mathematica package. Such routines can handle the singularities on the integration path which appear in the formulas. The stability of the numerical results has been tested against different choices of the parameters of the algorithm.

In an attempt to stay close to the notation of [2], we denote by Re($C^{\gamma}$) and Im($C^{\gamma}$) the real and imaginary parts of the MDM form factor for an incoming photon. Similarly, Re($C^{Z}$) and Im($C^{Z}$) denote the respective quantities for an incoming $Z$. The results are shown in Fig.5, Fig.6, and Fig.7. In what follows we always assume that the aforementioned quantities can be *individually* extracted from $e^{+}e^{-}$ experiments, through the methods discussed in detail in [2]. In all our calculations $m_t \geq 130\ GeV$, so the top is expected to decay before it can form a bound state [26], and therefore non-perturbative QCD effects should be numerically not very important. Even so, the one-loop QCD vertex corrections (graph 7 in Fig.3 with $\gamma \leftrightarrow gluon$) dominate in general the contribution of all other graphs. This dominance is due to the fact that the strong coupling $\alpha_s$, although it is small enough for perturbation theory to be trusted, it is still significantly larger than both the electro-magnetic and the weak couplings, which multiply all other graphs. We used $\alpha_s = 0.1$, $\alpha = \frac{1}{128}$, $cos\theta_w = 0.885$, $M_w = 80.6\ GeV$, $M_z = 91.1\ GeV$ and $m_b = 4.5\ GeV$. In addition, in the computations we report, we always place ourselves sufficiently above the respective $t\bar{t}$ threshold, so that non-perturbative threshold effects may also be omitted [27]. Of course, since the mass of the top is a free parameter which we vary in our computations, the location of the threshold ($s_{thr} = 4m_t^2$) is also variable.

In Fig.5, $\text{Re}(C^\gamma)$, $\text{Im}(C^\gamma)$, $\text{Re}(C^Z)$, and $\text{Im}(C^Z)$ are plotted as function of $\sqrt{s}$ for $m_t = 130\ GeV$ and two different values for the mass of the Higgs: a light one ($m_H = 150 GeV$) and a heavy one ($m_H = 500 GeV$). In Fig.6, $\text{Re}(C^\gamma)$, $\text{Im}(C^\gamma)$, $\text{Re}(C^Z)$, and $\text{Im}(C^Z)$ are plotted as function of $\sqrt{s}$ with a heavier choice for the top mass, namely $m_t = 200\ GeV$, and the same values for the Higgs masses as in Fig.5. Finally, in Fig.7 $\text{Re}(C^\gamma)$, $\text{Im}(C^\gamma)$, $\text{Re}(C^Z)$, and $\text{Im}(C^Z)$ are plotted as a function of $m_H$, for two different values of the top mass, $m_t = 130\ GeV$ and $m_t = 200\ GeV$. We fixed the value of $\sqrt{s}$ at 0.5 $TeV$, exactly as in the analysis of [2]. It is important to notice that in all cases shown in Fig.5 and Fig.6 the form factors approach zero, as $\sqrt{s}$ increases, at least for the range of the parameters that we have explored.

As we see in Fig.5, both $\text{Re}(C^\gamma)$ and $\text{Im}(C^\gamma)$ are smooth functions of $\sqrt{s}$ slightly above $\sqrt{s} = 300\ GeV$ for both values of $m_H$. On the other hand both $\text{Re}(C^Z)$ and $\text{Im}(C^Z)$ display a spike (peak) at around $\sqrt{s} = 600\ GeV$ when $m_H = 500\ GeV$, whereas they are smooth and have no spikes when $m_H = 150\ GeV$. Exactly the same qualitative behavior is also displayed by the graphs of Fig.6; the only difference is an overall enhancement in the values of all quantities shown, since the top is now heavier. Finally, in Fig.7 both $\text{Re}(C^\gamma)$ and $\text{Im}(C^\gamma)$ are smooth functions of $m_H$; on the contrary, $\text{Re}(C^Z)$ and $\text{Im}(C^Z)$ are rather sensitive to the values of $m_H$, especially if the mass of the top is large.

The Feynman graph responsible for the behavior described above is the non-Abelian graph denoted by "special" in Fig.3 (and its mirror graph, not shown), which exists only if the incoming gauge boson is a $Z$. On the other hand, the Abelian graph in Fig 3.10, common to both incoming $\gamma$ and $Z$, is of no particular importance. The importance of the "special" graph is due to the appearance of a threshold effect, as soon as the momentum $\sqrt{s}$, carried by the incoming $Z$, reaches the critical value $\sqrt{s_c}$, given by $\sqrt{s_c} = M_z + m_H$, namely the threshold for $Z + H$ production ($e^+e^- \rightarrow Z \rightarrow Z + H$) [28]. It is now easy to understand the behavior of Re($C^Z$) and Im($C^Z$) in Fig.5(c,d) and Fig.6(c,d): in the first case ($m_t = 130\ GeV$), for $m_H = 150\ GeV$ we have $\sqrt{s_c} = 241\ GeV$, which is *below* the $t\bar{t}$ production threshold $\sqrt{s_{thr}} = 260\ GeV$. So, the threshold effect of the "special" graph does not show up in the plots. On the contrary, for $m_H = 500\ GeV$, we have that $\sqrt{s_c} = 591\ GeV$, which gives the effect we observe in our plots, for both Re($C^Z$) and Im($C^Z$). Exactly the same happens in the second case ($m_t = 200\ GeV$), since the value of $\sqrt{s_c} = 591\ GeV$ is still larger than $\sqrt{s_{thr}} = 400\ GeV$. Of course, the value of $\sqrt{s_c}$ does not depend on the choices of $m_t$; what depends on the value of $m_t$ is whether or not the $HZ$-threshold effect at $\sqrt{s_c}$ appears in the plots.

The threshold effect seen in both Fig.7c and Fig.7d is also due to the "special" graph. For a given value of $\sqrt{s}$ we expect a threshold effect as soon as $m_H = m_H|_{thr} = \sqrt{s} - M_z$, thus, for incoming $\sqrt{s} = 500\ GeV$ one expects an effect at $m_H|_{thr} = 500\ GeV - M_z \approx 409\ GeV$; this is indeed what we observe in Fig.7c and Fig.7d. In particular, the position of $m_H^{thr}$ remains unchanged as it should, for both values of $m_t$ we examined. We emphasize that due to such threshold effect, some of our gauge-invariant form factors are much more sensitive to the value of the Higgs mass than the cross section as a whole [18]. For example, $Re(C^Z)$ of Fig.7c, for $m_t = 200\ GeV$ (dark curve), changes from Re($C^Z$) $= -1.4 \times 10^{-3}$ at $m_H = 200\ GeV$ to Re($C^Z$) $= -3.9 \times 10^{-3}$ at $m_H = 400\ GeV$, a change of $\approx 280\%$.

Based on what we said above about the $ZH$-threshold effect of the "special" graph, the following possibilities may come up, if one measured Re($C^Z$) and Im($C^Z$) in the future ($s_{max}$ is the maximum energy of the collider):

a) If $2m_t - M_z < m_H < \sqrt{s_{max}}$, a threshold effect should appear in *both* Re($C^Z$) and Im($C^Z$) when plotted as a function of $\sqrt{s}$, at a position $\sqrt{s_c} = m_H + M_z$.

b) If $m_H < 2m_t - M_z$ or $m_H > \sqrt{s_{max}}$, then no threshold effect should appear in either Re($C^Z$) or Im($C^Z$, in the context of the SM. The second case is particularly interesting, especially in view of the current theoretical bias toward a relatively "light" Higgs. For example if $m_H < 150\ \ GeV$, as favored by the Minimal Supersymmetric Standard Model, one should not see any $HZ$ threshold effects, unless it turns out that $m_t < 120\ \ GeV$ [29].

c) If a threshold effect appears in one of $\mathrm{Re}(C^Z)$ or $\mathrm{Im}(C^Z)$ but *not* in the other, or if more than one threshold effect appears in either or both of them, this should be interpreted as a signal of physics outside the SM.

It is important to keep in mind that if one calculates the form factors in a gauge-dependent way, i.e. without pinching, in some cases one gets wrong predictions of threshold effects, related to unphysical, gauge-dependent thresholds. In these cases, the difference between the gauge-independent and the gauge-dependent prediction becomes, of course, dramatic.

## 7. Conclusions

In this paper we addressed some of the theoretical issues involved in the computation of off-shell form factors in the context of the Standard Model. Such form factors can be extracted, at least in principle, from cross-sections in future $e^+e^-$ experiments. In particular, we explicitly demonstrated that the conventionally defined MDM form factors of the top quark are gauge-dependent in the class of $R_\xi$ gauges. Due to the massiveness of the top quark, this gauge dependence turned out to be numerically very strong. We used the S-matrix pinch technique to eliminate the gauge dependence, and defined a gauge-independent MDM form factor, suitable for comparison with experiment. We proved a simple QED-like Ward identity, satisfied by the gauge-invariant $\gamma t\bar{t}$ vertex costructed via the PT. Finally, we computed the gauge-independent MDM form factors and presented the results for both real and imaginary (absorptive) parts. A very interesting feature of such form factors is their high sensitivity to the value of the Higgs mass, which enters the formulas as a parameter. For this reason, assuming that the top quark will be discovered and detected before the Higgs particle, the form factor we have analyzed in the present paper contains interesting information about the Higgs boson.

Of particular interest for the study of CP violation will be the experimental measurement of EDM form factors in the future. As we mentioned in the introduction, there are no EDM contributions within the SM up to three loops, but several models predict non-zero EDM already at one-loop. We are currently investigating issues of gauge invariance in the context of some of those models, with focus on the EDM form factors, and will present our results elsewhere.

## 8. Acknowledgements

The authors thank Professor A. Sirlin for useful discussions. One of us (C.P.) thanks the Physics Department of the New York University for its hospitality during the early stages of this work. This work was supported in part by the National Science Foundation under Grant No.PHY-9017585.

## 9. Appendix

In this Appendix we list all the contributions to the magnetic form factor $F(q^2)$ of the top quark. Using the standard Feynman parametrization formula

$$\frac{1}{ABC} = \int_0^1 dx \int_0^{1-x} dy \frac{1}{[Ax + By + C(1-x-y)]^3} \tag{9.1}$$

and subsequently the convenient change of variables $t = x + y, \quad tz = x - y$, we can cast the MDM contributions of the graphs in Fig.3 (except the one denoted "special") in the following form

$$F(q^2) = \frac{\alpha}{2\pi} \left\{ m_t^2 \sum_i N_{(i)}^V \int_0^1 dt\ t\ \left[ A_{(i)}^V + B_{(i)}^V t + C_{(i)}^V t^2 \right] \int_0^1 dz\ L_{(i)}(t,z) \right\} \tag{9.2}$$

where $L_{(i)}(t,z) = L(t,z;m_1,m_2,q^2)$.

In what follows we will use the short-hand notation $L(m_1,m_2)$. In order to identify the contributions to the magnetic form factor $F^\gamma(q^2)$ and $F^Z(q^2)$, which arise from the $\gamma t\bar{t}$ and the $Zt\bar{t}$ vertices respectively, the coefficients in Eq. (9.2) carry a vertex label $V = \gamma,\ Z$. Obviously the integrand functions $L_{(i)}(z)$ do not have such a label, as they do not depend on which gauge boson enters the vertex. In the following list, the contributions are listed in the same order as in Fig.3 and we use again the abbreviations $s = sin\theta_W,\ c = cos\theta_W$ and $\kappa = |U_{tb}|^2$. In addition, we define $\mathcal{L} = (1 - \frac{8}{3}s^2)$.

$$N^\gamma_{(1+2+3)} = \frac{3\kappa}{4s^2} \qquad A^\gamma_{(1+2+3)} = 0 \qquad B^\gamma_{(1+2+3)} = 1 \qquad C^\gamma_{(1+2+3)} = 1$$
$$L_{(1+2+3)} = L(M_w^2, m_b^2) \tag{9.3}$$

$$N^\gamma_{(4)} = \frac{3\kappa}{8\ M_w^2 s^2} \qquad A^\gamma_{(4)} = 2m_b^2 \qquad B^\gamma_{(4)} = -3m_b^2 - m_t^2 \qquad C^\gamma_{(4)} = m_b^2 + m_t^2$$
$$N^Z_{(4)} = \frac{3\kappa}{8\ M_w^2 s^2}\left(\frac{c}{s} - \frac{1}{2s\ c}\right) \qquad A^Z_{(4)} = 2m_b^2 \qquad B^Z_{(4)} = -3m_b^2 - m_t^2 \qquad C^Z_{(4)} = m_b^2 + m_t^2$$
$$L_{(4)} = L(M_w^2, m_b^2) \tag{9.4}$$

$$N^\gamma_{(5)} = \frac{\kappa}{4\ s^2} \qquad A^\gamma_{(5)} = 2 \qquad B^\gamma_{(5)} = -3 \qquad C^\gamma_{(5)} = 1$$
$$N^Z_{(5)} = -\frac{3\kappa}{8\ s^3 c} \qquad A^Z_{(5)} = -2 + \frac{4}{3}s^2 \qquad B^Z_{(5)} = 3 - 2s^2 \qquad C^Z_{(5)} = -1 + \frac{2}{3}s^2$$
$$L_{(5)} = L(m_b^2, M_w^2) \tag{9.5}$$

$$N^\gamma_{(6)} = -\frac{1}{4\ s^2 c^2} \qquad A^\gamma_{(6)} = 2 \qquad B^\gamma_{(6)} = -3 - \frac{32}{9}s^4 + \frac{8}{3}s^2 \qquad C^\gamma_{(6)} = \mathcal{L} + \frac{32}{9}s^4$$

$$N^Z_{(6)} = -\frac{3}{64\ s^3 c^3} \qquad A^Z_{(6)} = 8\ \mathcal{L} \qquad B^Z_{(6)} = -\mathcal{L}(\mathcal{L}^2+11) \qquad C^Z_{(6)} = \mathcal{L}(\mathcal{L}^2+3)$$

$$L_{(6)} = L(m_t^2, M_z^2) \tag{9.6}$$

$$N^\gamma_{(7)} = -\frac{8}{9} \qquad A^\gamma_{(7)} = 0 \qquad B^\gamma_{(7)} = -1 \qquad C^\gamma_{(7)} = 1$$

$$N^Z_{(7)} = -\frac{1}{3\ sc}\ \mathcal{L} \qquad A^Z_{(7)} = 0 \qquad B^Z_{(7)} = -1 \qquad C^Z_{(7)} = 1$$

$$L_{(7)} = L(m_t^2, 0) \tag{9.7}$$

$$N^\gamma_{(8)} = \frac{\kappa}{8\ M_w^2 s^2} \qquad A^\gamma_{(8)} = 0 \qquad B^\gamma_{(8)} = m_b^2 - m_t^2 \qquad C^\gamma_{(8)} = m_b^2 + m_t^2$$

$$N^Z_{(8)} = -\frac{3\kappa}{64\ M_w^2 s^3 c} \qquad A^Z_{(8)} = 0 B^Z_{(8)} = 4(1-\frac{2}{3}s^2)(m_t^2 - m_b^2) \qquad C^Z_{(8)} = -4m_t^2(1-\frac{2}{3}s^2) + \frac{8}{3}m_b^2 s^2$$

$$L_{(8)} = L(m_b^2, M_w^2) \tag{9.8}$$

$$N^\gamma_{(9)} = -\frac{m_t^2}{4M_w^2 s^2} \qquad A^{\gamma)}_{(9)} = 0 \qquad B^\gamma_{(9)} = 0 \qquad C^\gamma_{(9)} = 1$$

$$N^Z_{(9)} = -\frac{3\ m_t^2}{32\ M_w^2 s^3 c}\mathcal{L} \qquad A^Z_{(9)} = 0 \qquad B^Z_{(9)} = 0 \qquad C^Z_{(9)} = 1$$

$$L_{(9)} = L(m_t^2, M_z^2) \tag{9.9}$$

$$N^\gamma_{(10)} = \frac{m_t^2}{4\ M_W^2 s^2} \qquad A^\gamma_{(10)} = 0 \qquad B^\gamma_{(10)} = 2 \qquad C^\gamma_{(10)} = -1$$

$$N^Z_{(10)} = \frac{3\ m_t^2}{32\ M_W^2 s^3 c}\mathcal{L} \qquad A^Z_{(10)} = 0 \qquad B^Z_{(10)} = 2 \qquad C^Z_{(10)} = -1$$

$$N^Z_{(10)} = \frac{3\ m_t^2}{32\ M_W^2 s^3 c}\mathcal{L} \qquad A^Z_{(10)} = 0 \qquad B^Z_{(10)} = 2 \qquad C^Z_{(10)} = -1$$

$$L_{(10)} = L(m_t^2, M_H^2) \tag{9.10}$$

For the "special" graph (last one in Fig.3) and its mirror graph (not shown) the original Feynman parametrization given in Eq 9.1 was maintained, since the change of variables $(x, y) \to (t, z)$, used for all other graphs, is not convenient.

Finally, the one-loop QCD contribution is obtained from Eq. 9.6, by multiplying by the factor $\frac{\alpha_s}{2\alpha}$, where $\alpha_s$ is the QCD coupling at $\mu = M_z$, namely

$$\begin{aligned} N^{\gamma}_{QCD} &= [\frac{\alpha_s}{2\alpha}]N^{\gamma}_{(7)} \\ N^{Z}_{QCD} &= [\frac{\alpha_s}{2\alpha}]N^{Z}_{(7)} \end{aligned} \tag{9.11}$$

## 10. References.

29. Of course in such a case the usual arguments about the non-hadronization of the produced tops will be invalidated, and non-perturbative QCD effects may become important.

## 11. Figure Captions

1) Graphs (a)-(c) are some of the contributions to the S-matrix $T$. Graphs (e) and (f) are pinch parts, which, when added to the usual self-energy graphs (d), give rise to a gauge-independent effective self-energy.

2a) The two graphs defining the gauge-invariant vertex $\Gamma_{\mu}^{\alpha}$. The circle in the first graph indicates the $\Gamma_{\mu}^{F}$ part of the three-gluon vertex.

2b) Vertex-like pinch contributions from a box graph, in a general $R_{\xi}$ gauge, with $\xi \neq 1$.

3) All the graphs contributing to the $\gamma t\bar{t}$ and $Zt\bar{t}$ vertices.

4) The top quark MDM as a function of $\sqrt{s}$, computed for several different values of the gauge-fixing parameter $\xi_{w}$ (b,c,d,e), to be contrasted to the gauge-invariant answer shown in (a). The correspondance is as follows: $\xi_{w} = 10$ (b), $\xi_{w} = 100$ (c), $\xi_{w} = 1000$ (d), $\xi_{w} \to \infty$ (e).

5) The case of a "light" top ($m_{t} = 130\ GeV$), for two different choices of the Higgs boson mass: $m_{H} = 150\ GeV$, and $m_{H} = 500\ GeV$ (darker curve). In particular:

5a) Real part of MDM for an incoming $\gamma$,

5b) Imaginary part of MDM for an incoming $\gamma$,

5c) Real part of MDM for an incoming $Z$,

5d) Imaginary part of MDM for an incoming $Z$.

6) The case of a "heavy" top ($m_{t} = 200\ GeV$), for the same two choices of the Higgs boson mass as in Fig.5 :

6a) Real part of MDM for an incoming $\gamma$,

6b) Imaginary part of MDM for an incoming $\gamma$,

6c) Real part of MDM for an incoming $Z$,

6d) Imaginary part of MDM for an incoming $Z$.

7) The dependence of MDM on the mass $m_H$ of the Higgs boson, for a "light" and a "heavy" top, at $\sqrt{s} = 500\ GeV$:

7a) Real part of MDM for an incoming $\gamma$,

7b) Imaginary part of MDM for an incoming $\gamma$,

7c) Real part of MDM for an incoming $Z$,

7d) Imaginary part of MDM for an incoming $Z$.

# Gauge invariant top quark form factors from $e^{+}e^{-}$ experiments.

J. Papavassiliou[1] and C. Parrinello[2].

[1] Department of Physics, New York University, 4 Washington Place, New York, NY 10003, USA.

[2] Department of Physics, University of Edinburgh, Mayfield Road Edinburgh EH9 3JZ, Scotland, UK

ABSTRACT

Motivated by the possibility of experimental determination of top quark form factors in upcoming $e^+e^-$ experiments, as recently discussed by Atwood and Soni, we show at the one loop level that the conventionally defined magnetic dipole moment (MDM) form factor of the top quark is gauge dependent, in the class of renormalizable ($R_\xi$) gauges. We explicitly calculate its gauge dependence, which, due to the massiveness of the top quark, turns out to be numerically very sizable. We show how to use the S-matrix pinch technique in order to define gauge independent form factors. The real and imaginary parts of the gauge independent MDM form factor are calculated and their dependence on the top quark and Higgs boson masses is discussed. Most noticeably, the dependence on the Higgs mass turns out to be stronger than that of the cross section as a whole.

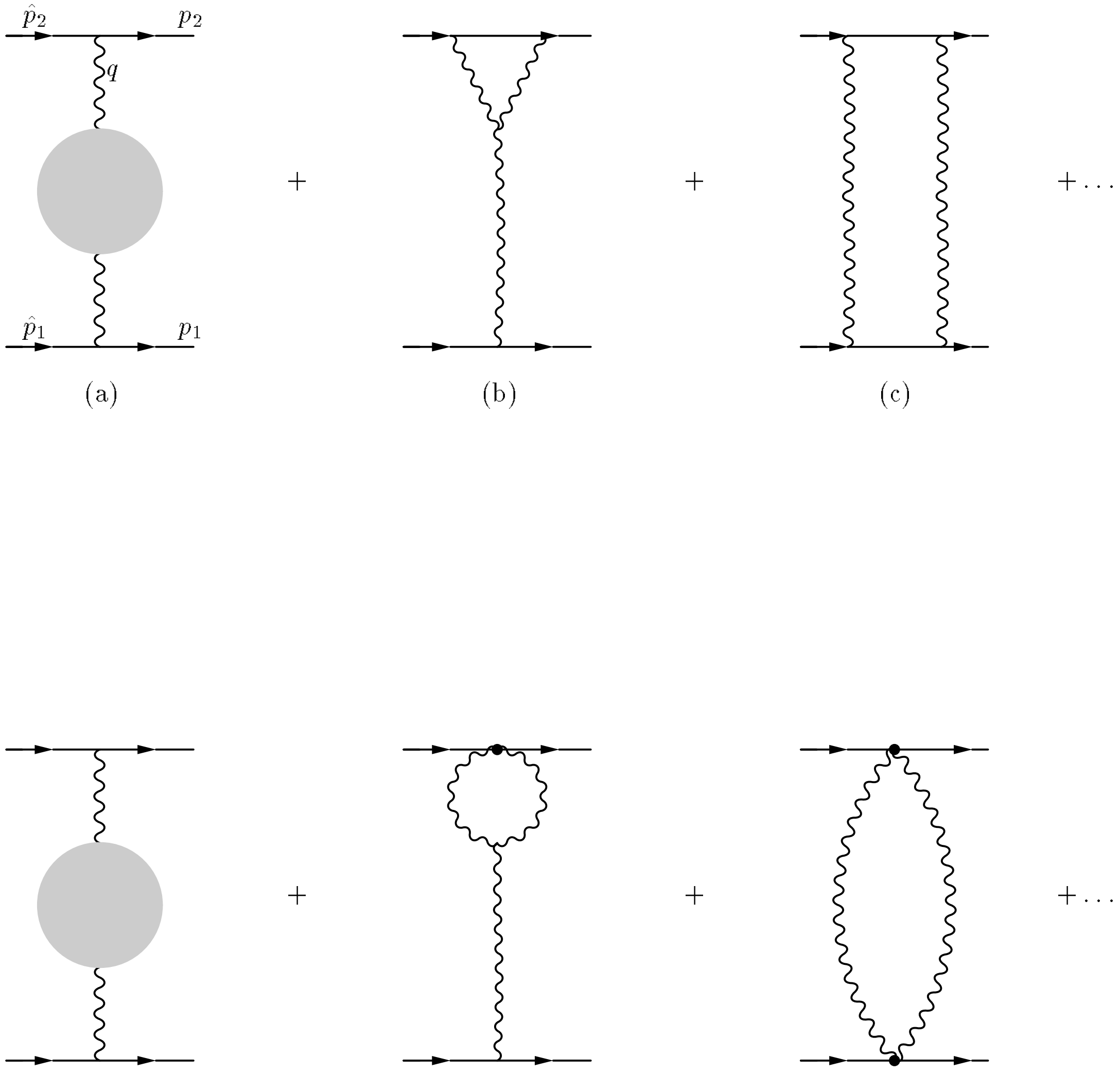

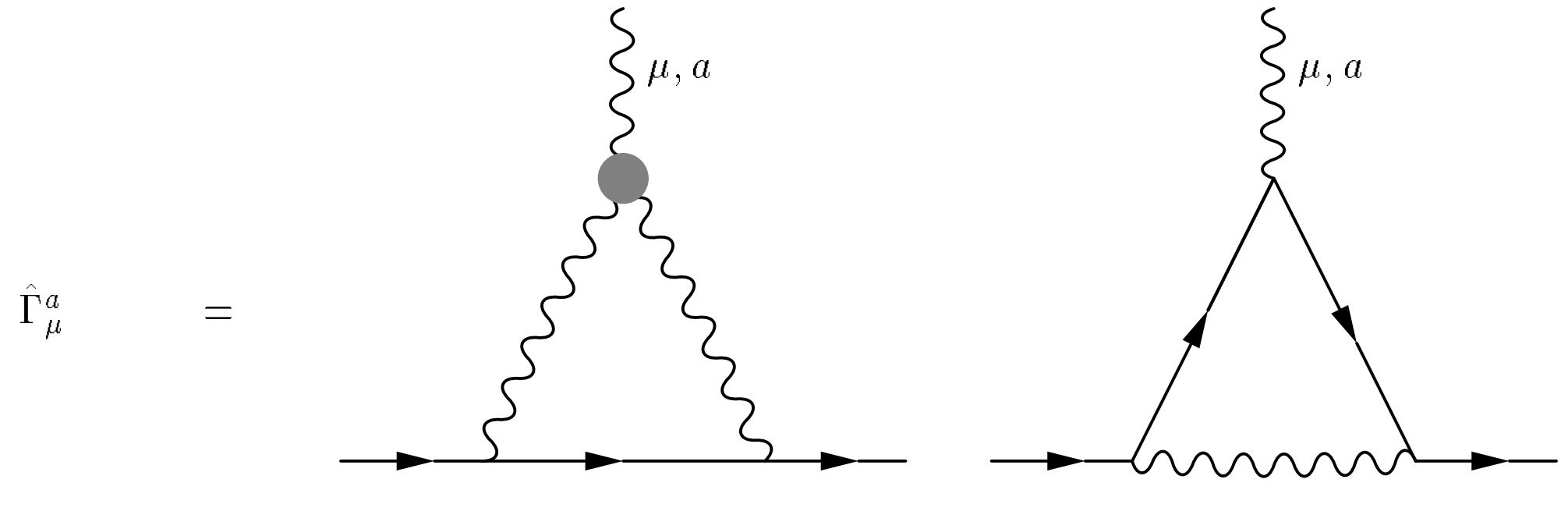

(2a)

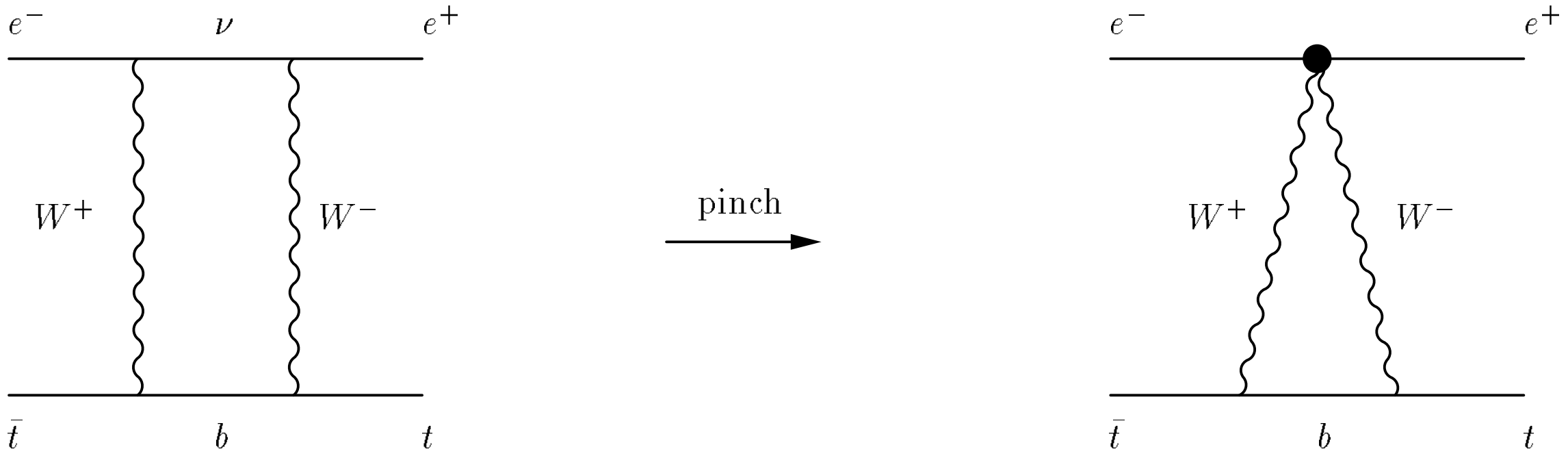

(2b)

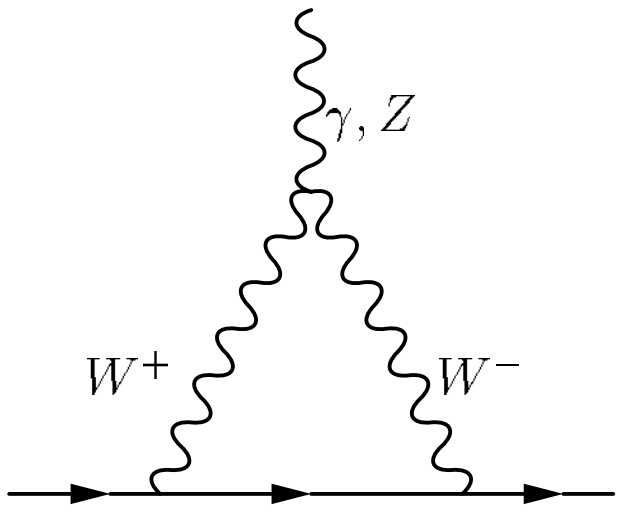

(1)

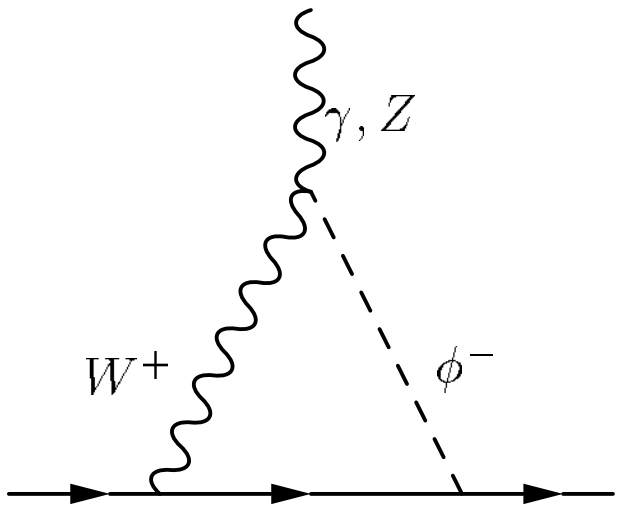

(2)

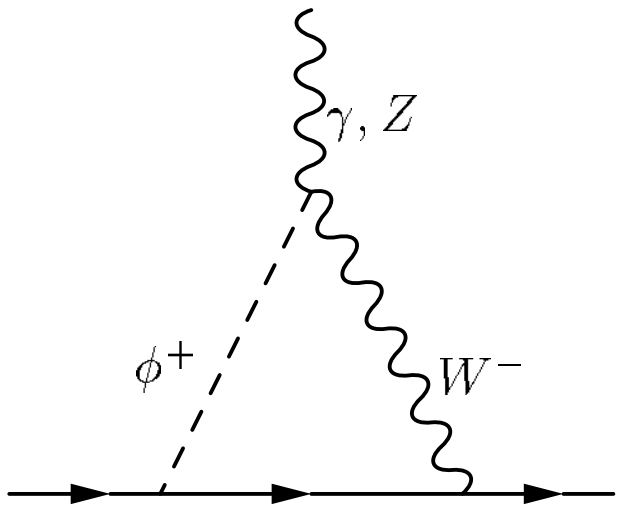

(3)

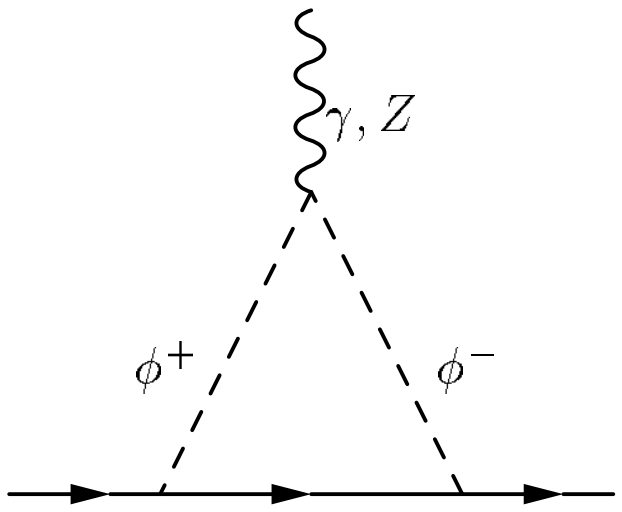

(4)

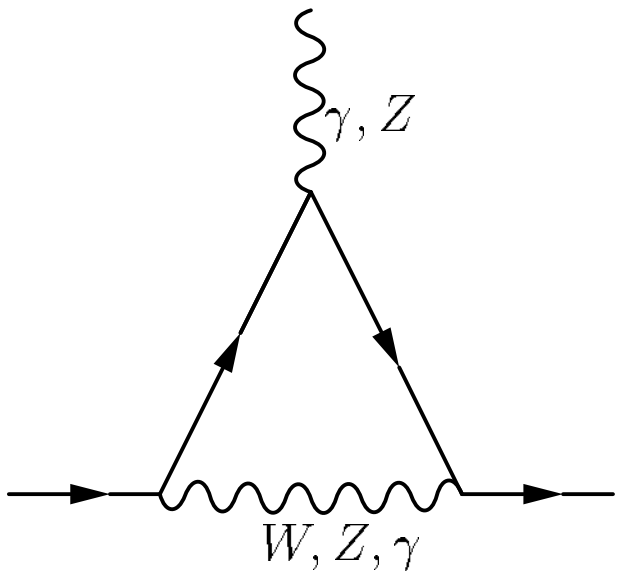

(5, 6, 7)

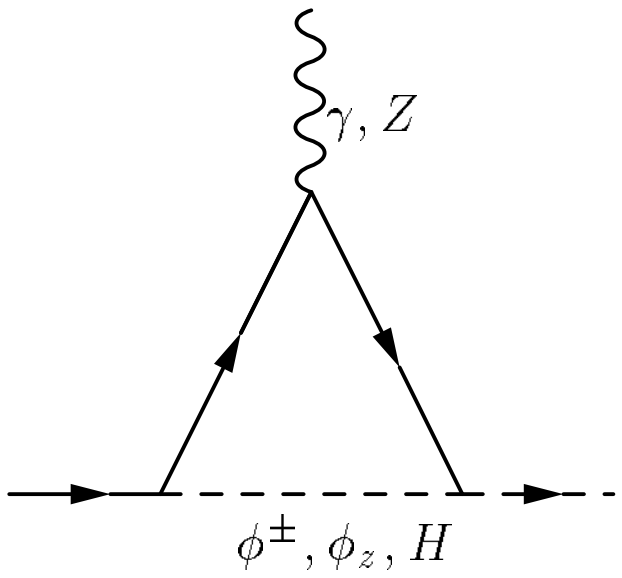

(8, 9, 10)

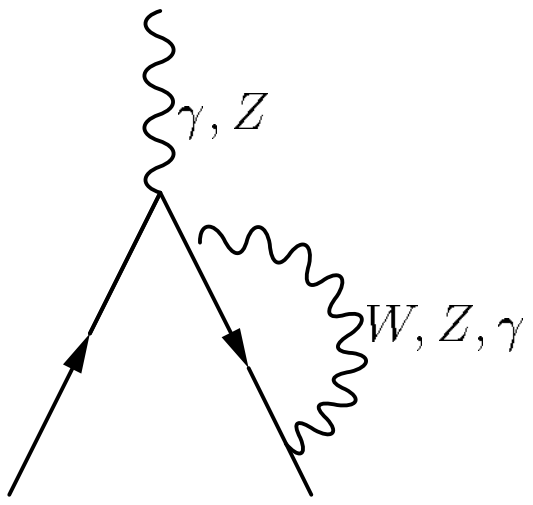

(11, 12, 13)

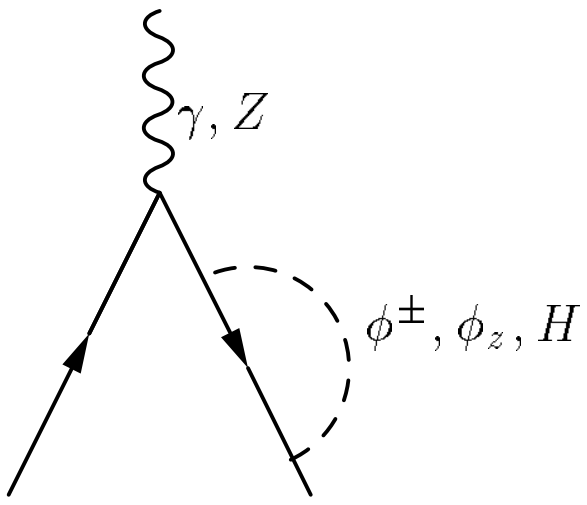

(14, 15, 16)

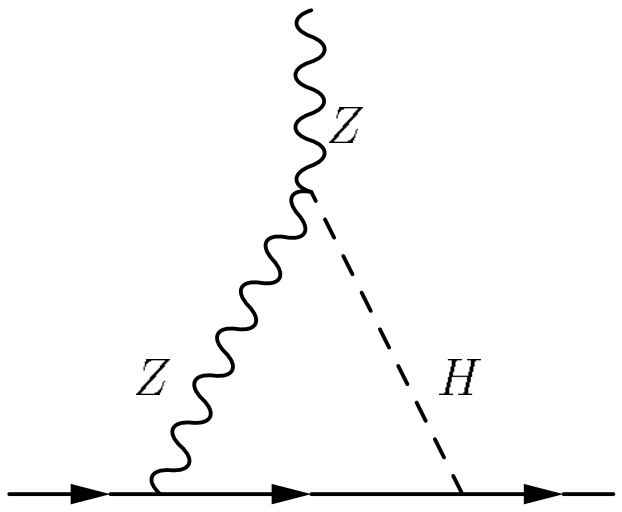

(*special*)

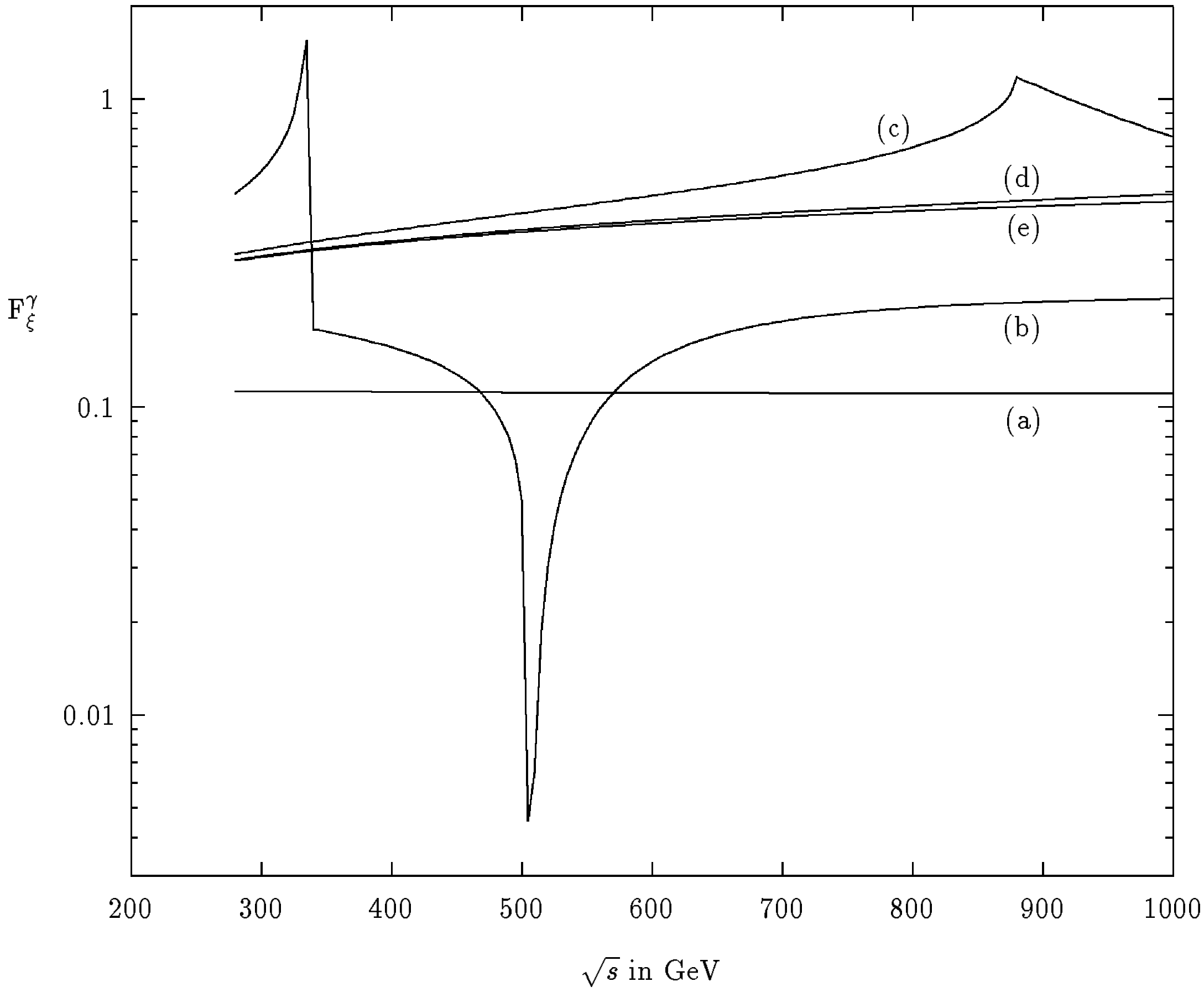

Fig. 4

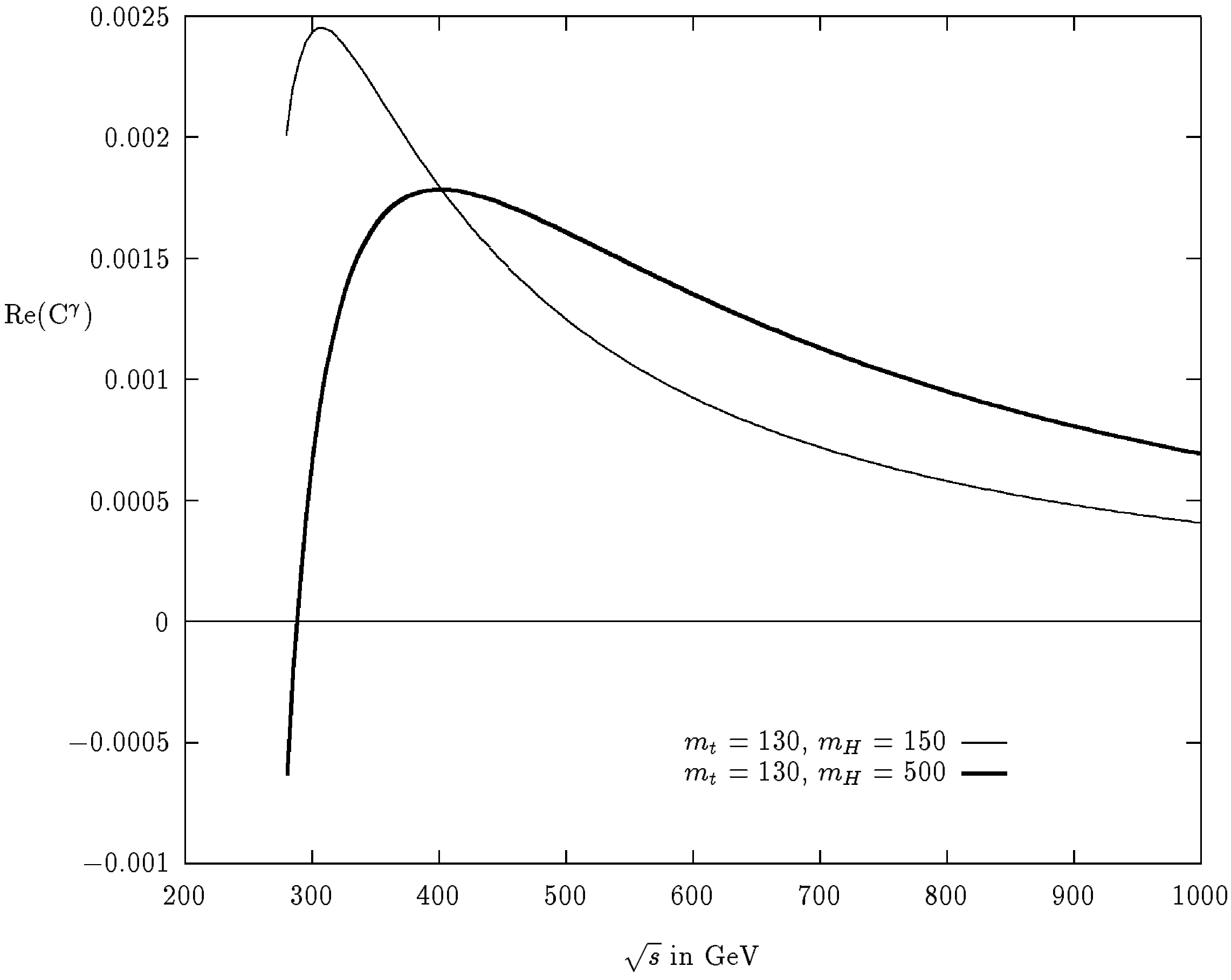

Fig. 5.a

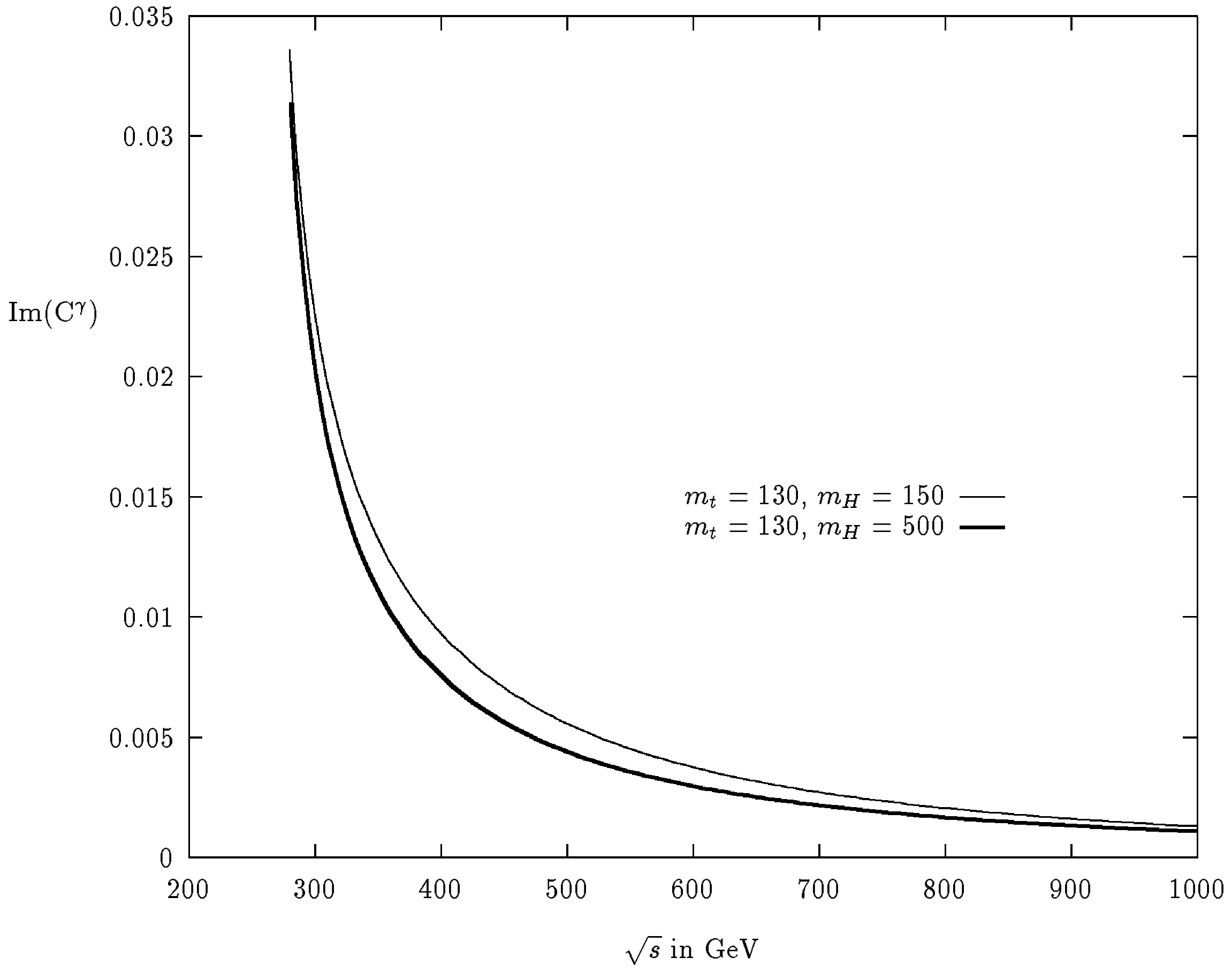

Fig. 5.b

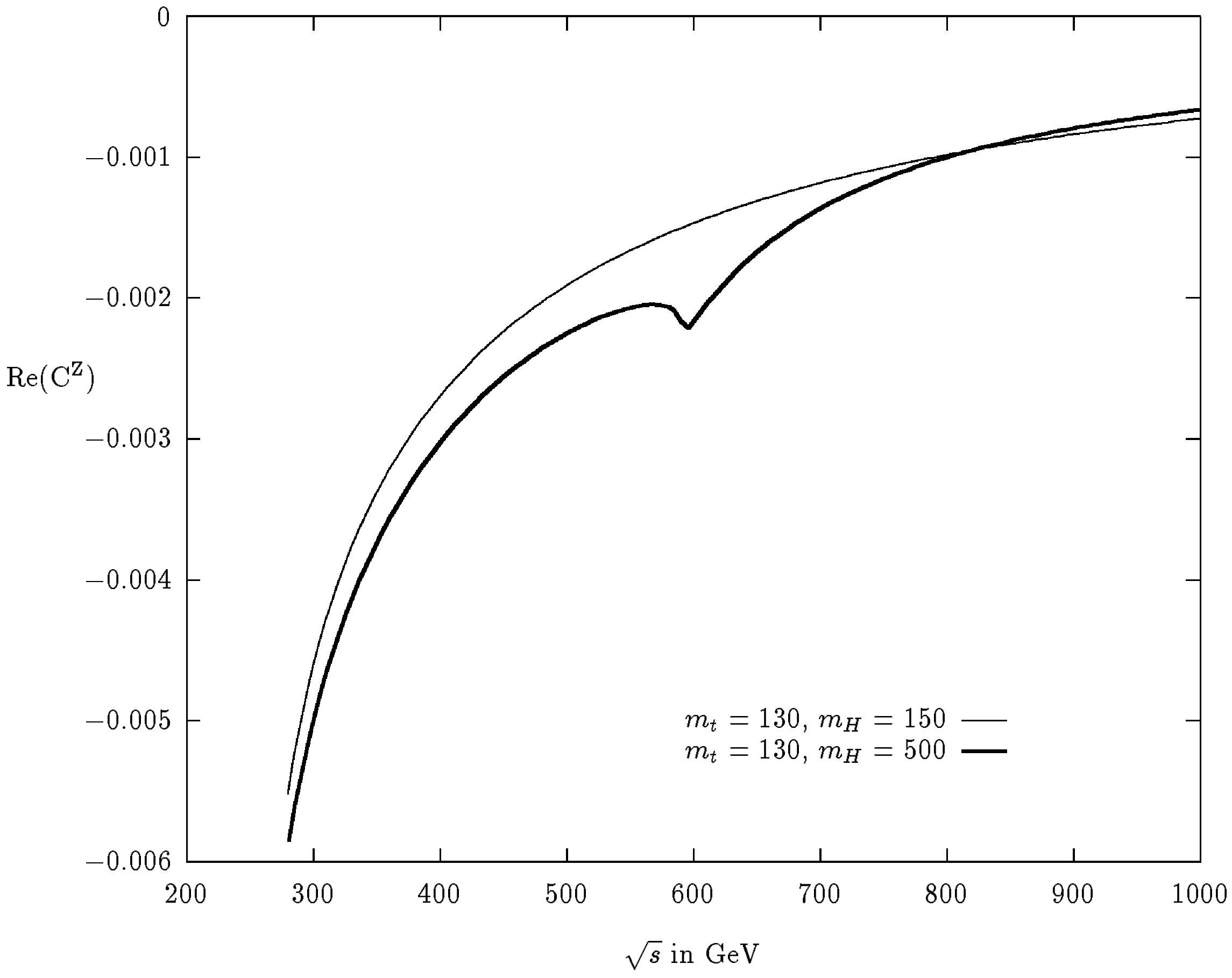

Fig. 5.c

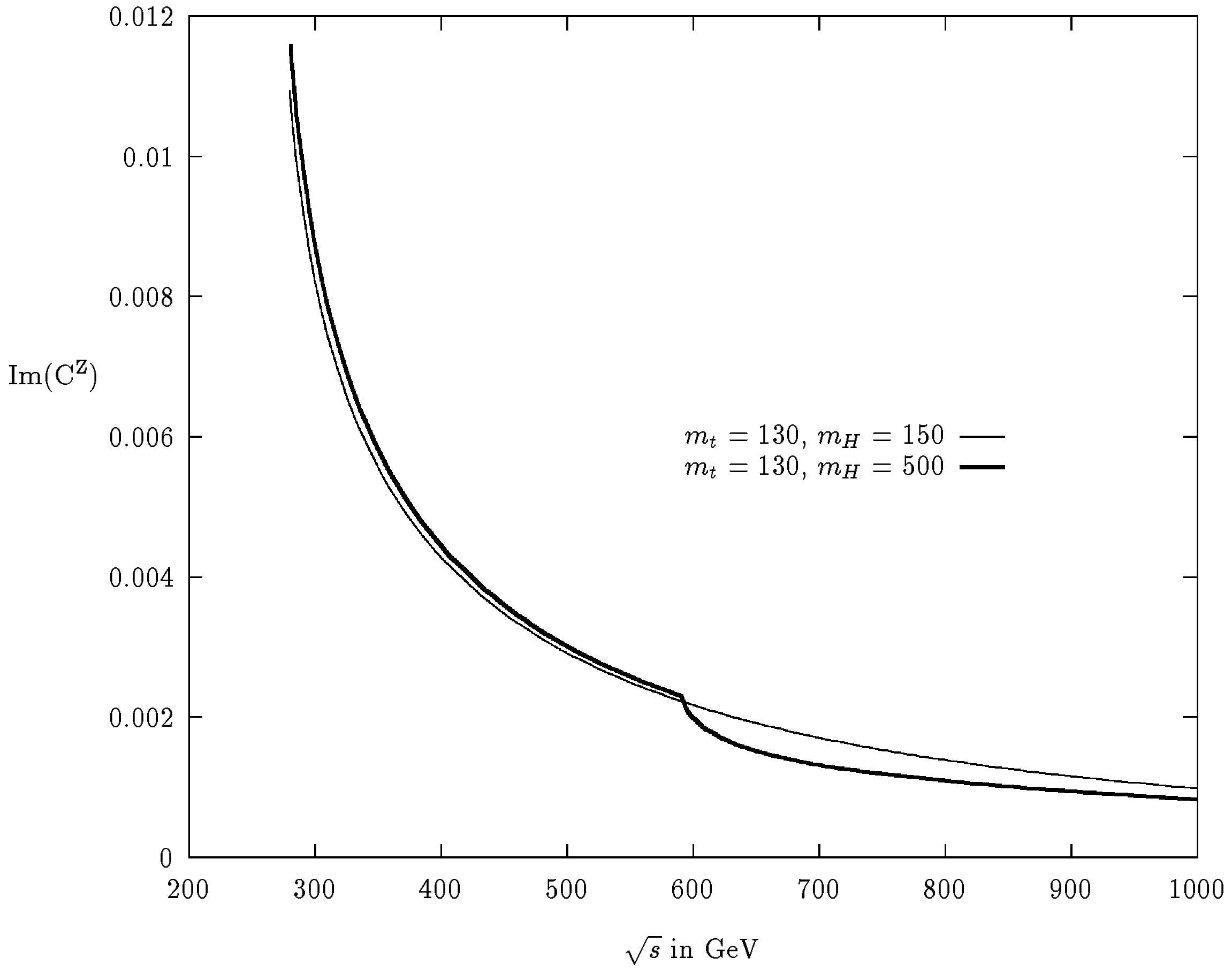

Fig. 5.d

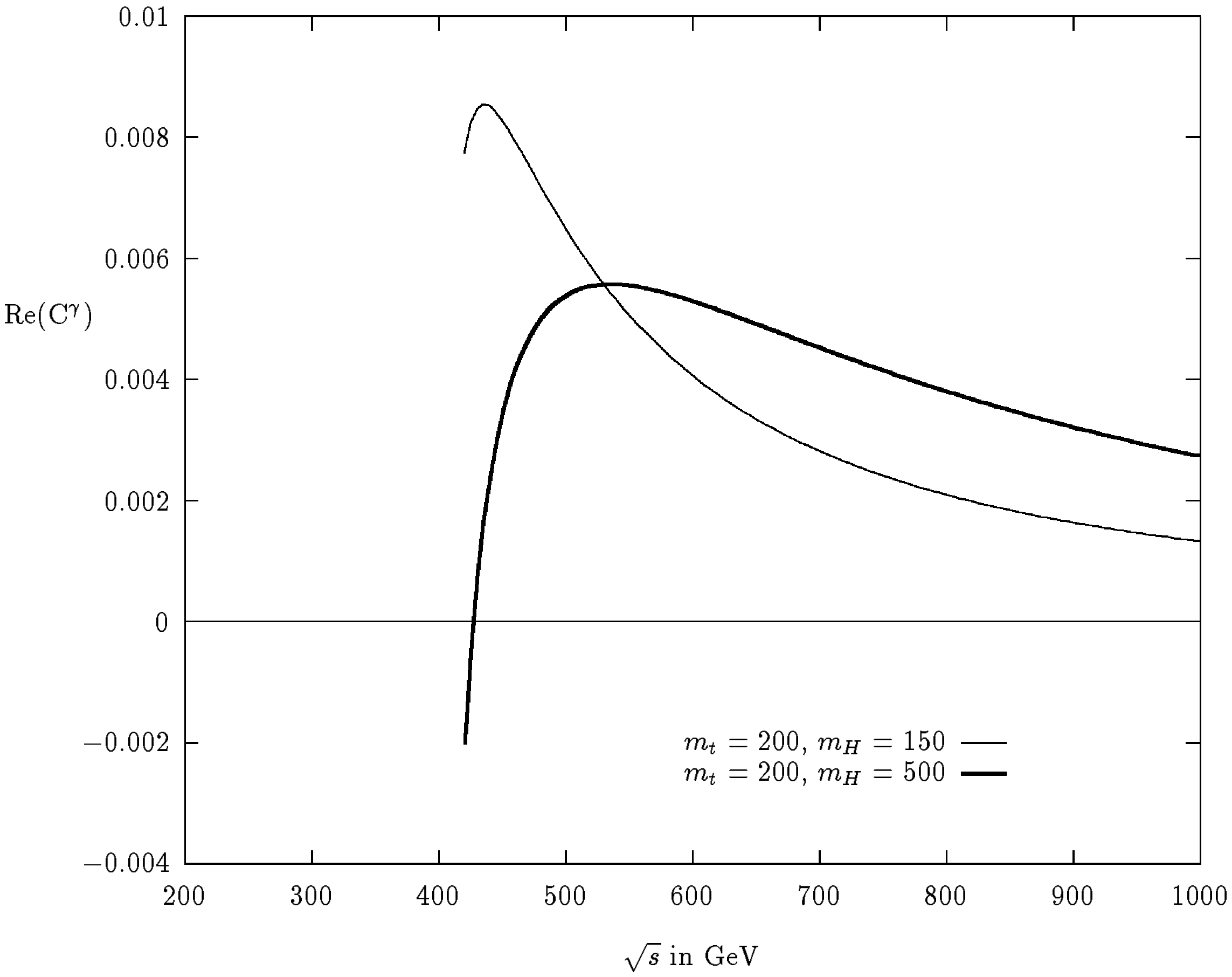

Fig. 6.a

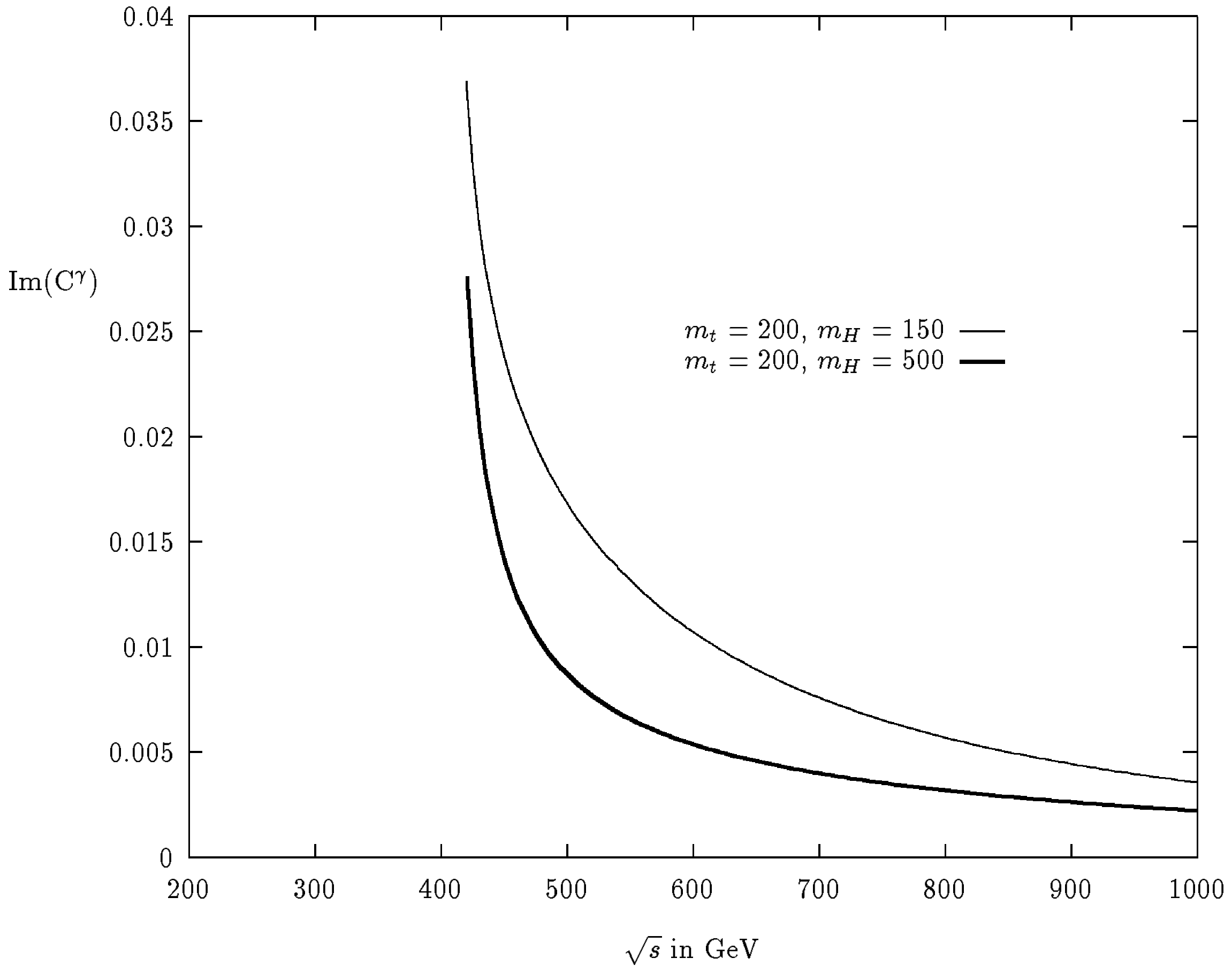

Fig. 6.b

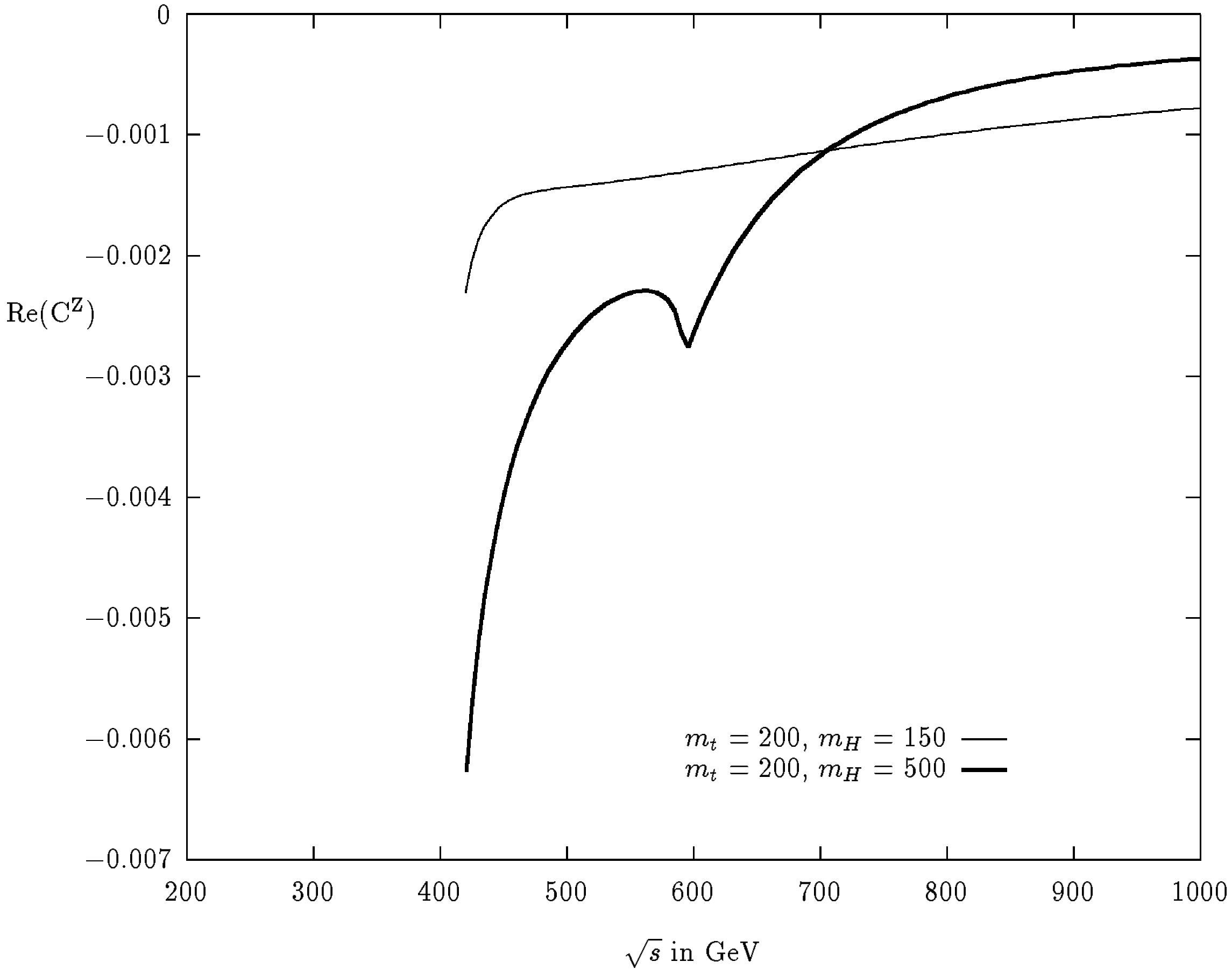

Fig. 6.c

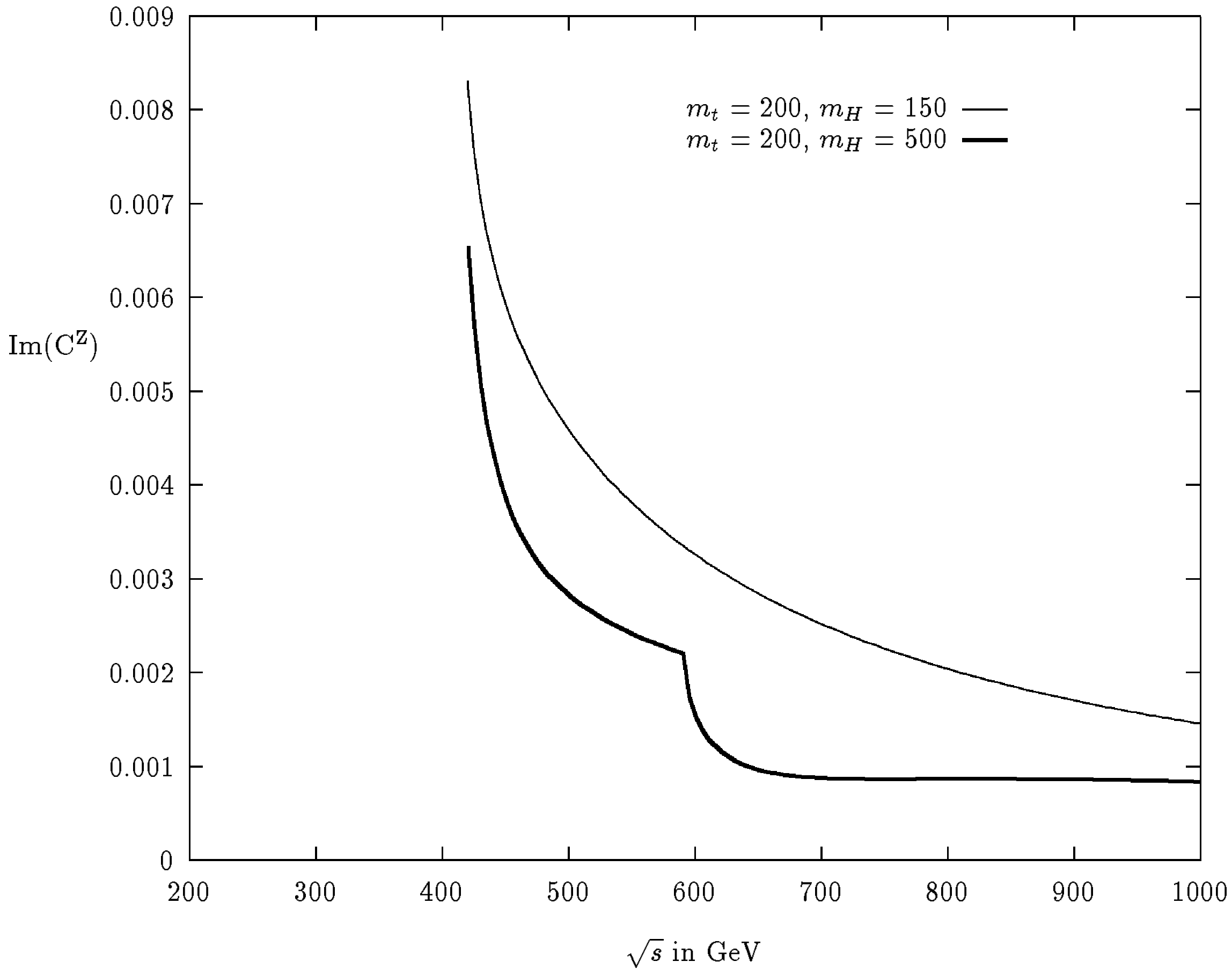

Fig. 6.d

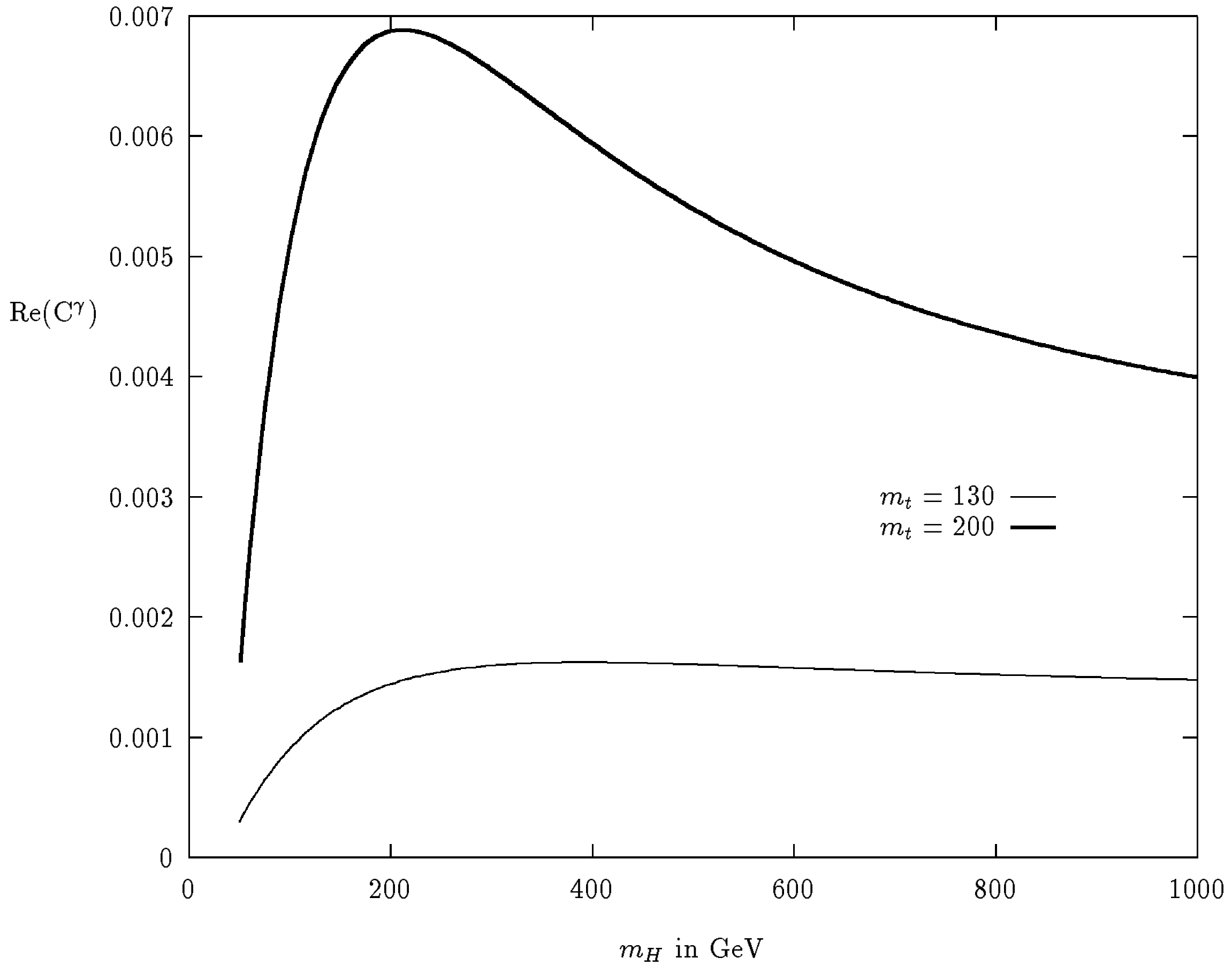

Fig. 7.a

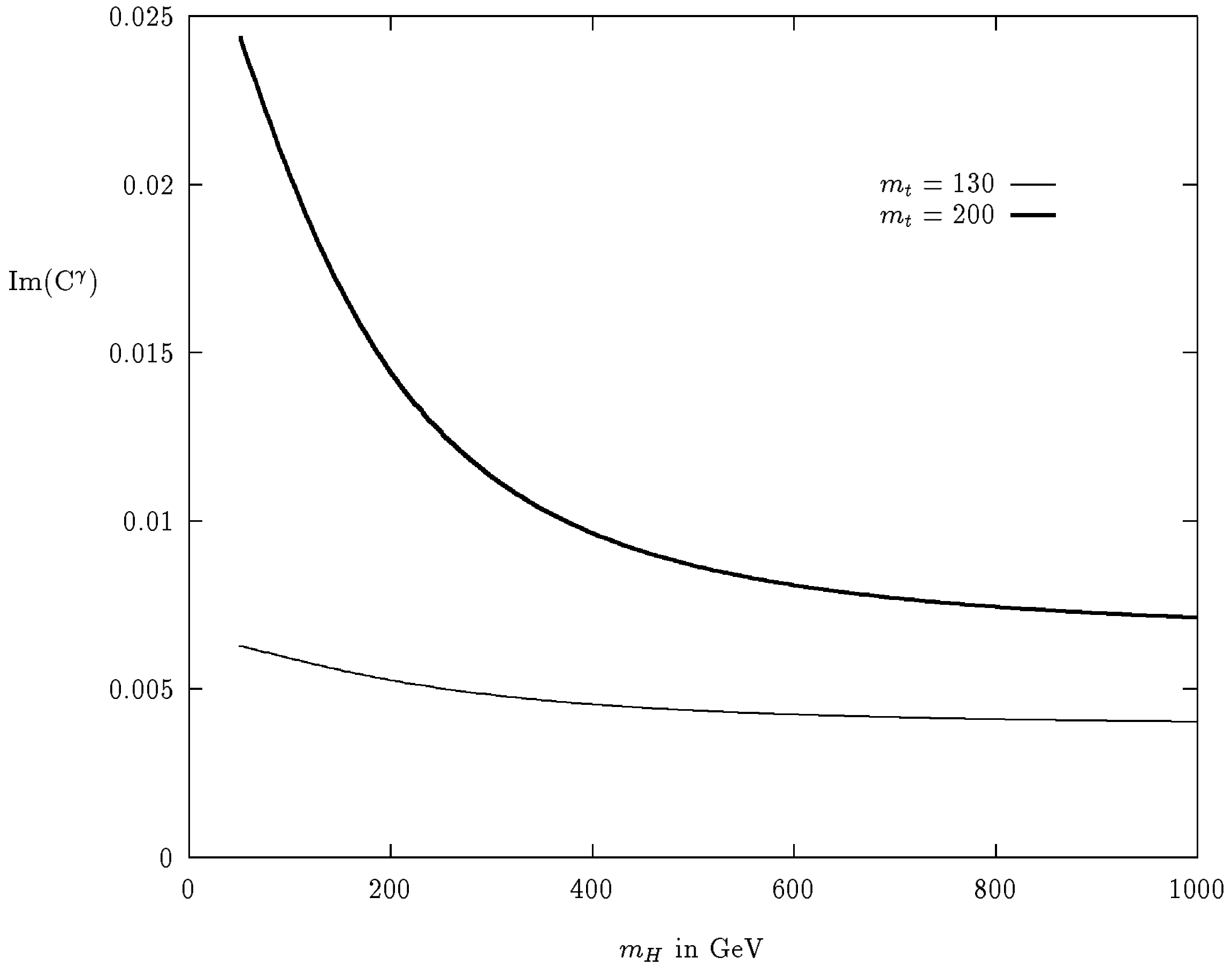

Fig. 7.b

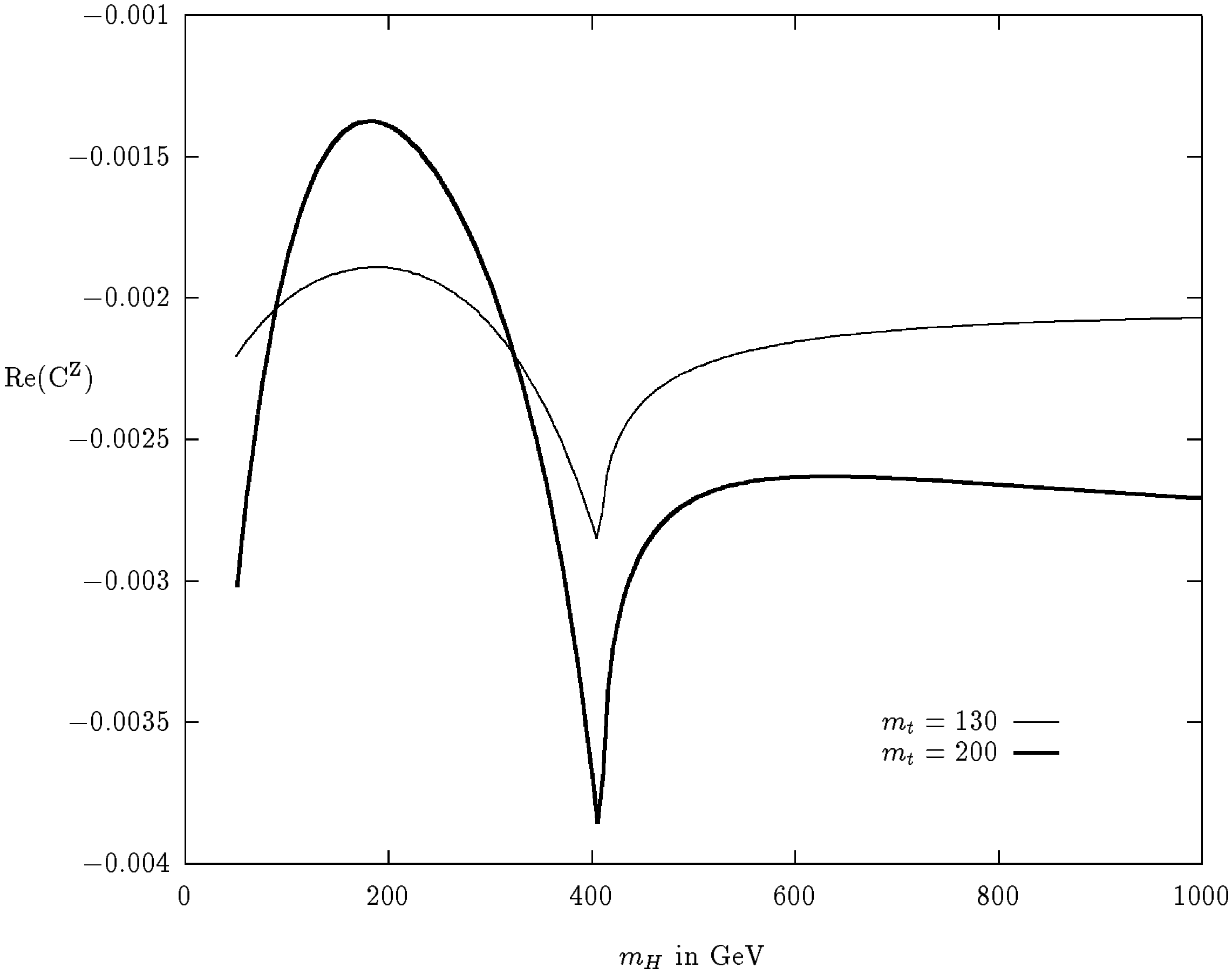

Fig. 7.c

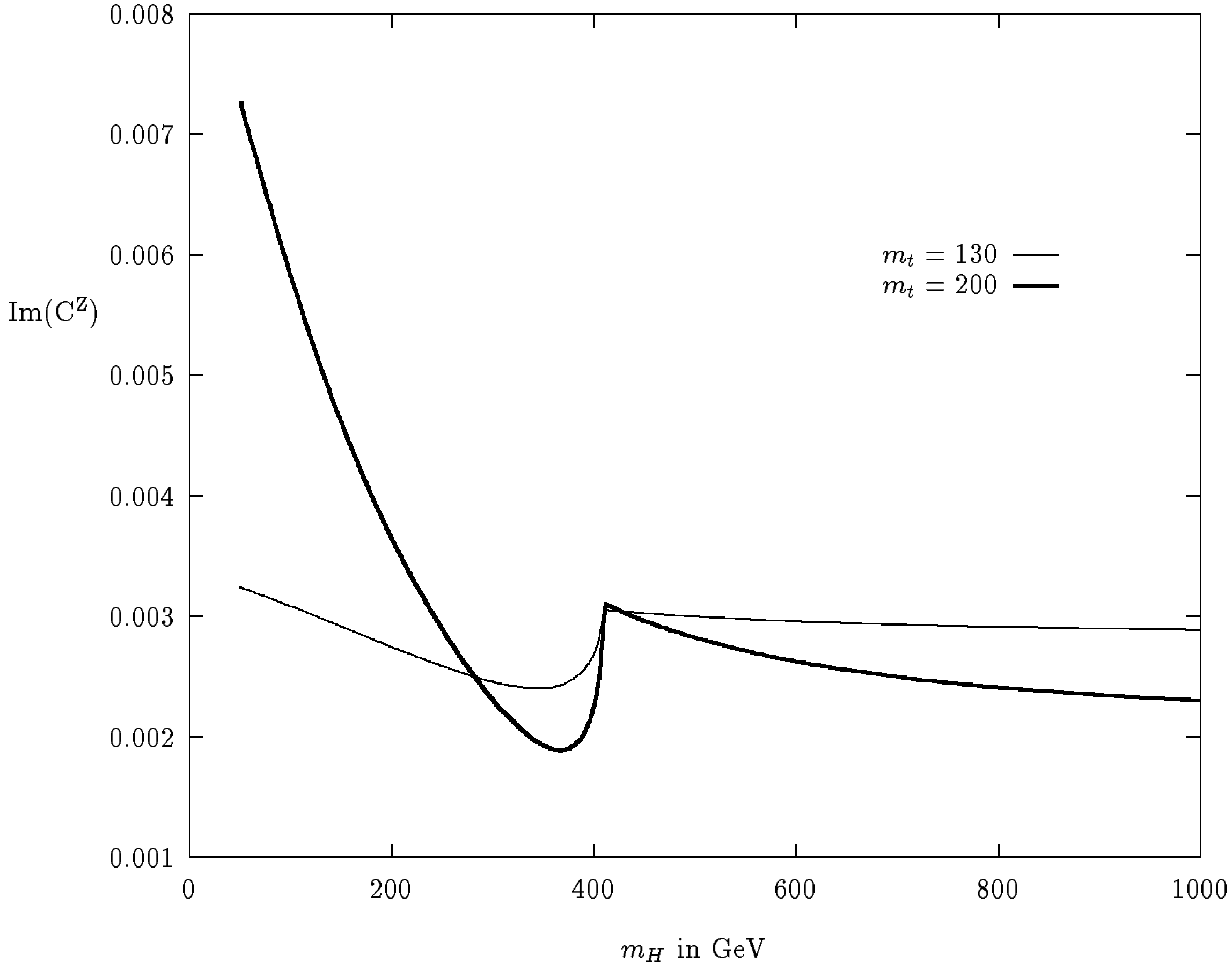

Fig. 7.d